\documentclass[twocolumn,dvipsnames,twocolappendix]{aastex631}
\usepackage{xspace}
\usepackage{amsmath}
\usepackage{lipsum, babel, ulem, soul, cancel}
\usepackage[shortlabels]{enumitem}
\usepackage{hhline}
\usepackage{multirow}

\newcommand{\paired}{\texttt{paired}\xspace}

\newcommand{\galah}{{\small GALAH}\xspace}
\newcommand{\apogee}{{\small APOGEE}\xspace}
\newcommand{\galex}{{\small GALEX}\xspace}
\newcommand{\wise}{{\small WISE}\xspace}
\newcommand{\lamost}{{\small LAMOST}\xspace}
\newcommand{\rave}{{\small RAVE}\xspace}

\newcommand{\gaia}{\textit{Gaia}\xspace}

\newcommand{\solmass}{M$_\odot$\xspace}

\newcommand{\app}{$\sim$}
\newcommand{\logg}{$\log g$\xspace}
\newcommand{\teff}{T$_\textrm{eff}$\xspace}
\newcommand{\feh}{[Fe/H]\xspace}
\newcommand{\mgfe}{[Mg/Fe]\xspace}

\newcommand{\ali}{$A\textrm{(Li)}$\xspace}

\newcommand{\halpha}{H-$\alpha$\xspace}
\newcommand{\vsini}{$v\sin i$\xspace}

\newcommand{\vbr}{$v_\textrm{broad}$\xspace}

\newcommand{\kms}{km s$^{-1}$\xspace}
\newcommand{\tc}{$T_c$\xspace}
\newcommand{\ang}{$\textrm{\AA}$\xspace}
\newcommand{\chisq}{$\chi^2$\xspace}

\newcommand{\li}{lithium\xspace}
\newcommand{\lirich}{Li-rich\xspace}
\newcommand{\lin}{Li-normal\xspace}

\newcommand{\dg}{doppelg\"anger\xspace}
\newcommand{\dgs}{doppelg\"angers\xspace}
\newcommand{\plen}{planetary engulfment\xspace}
\newcommand{\seismo}{asteroseismology\xspace}
\newcommand{\ber}{beryllium\xspace}
\newcommand{\cond}{condensation temperature\xspace}

\newcommand{\sprocess}{$s-$process\xspace}

\newcommand{\total}{1099\xspace}
\newcommand{\totaldg}{1099\xspace}
\newcommand{\totalrvs}{667\xspace}
\newcommand{\mstars}{25\xspace}
\newcommand{\wstars}{216\xspace}
\newcommand{\nbase}{71\xspace}
\newcommand{\nrgb}{136\xspace}
\newcommand{\nrc}{648\xspace}
\newcommand{\richgalex}{361\xspace}
\newcommand{\poorgalex}{289\xspace}

\definecolor{forestgreen}{rgb}{0.13, 0.55, 0.13}
\definecolor{neongreen}{rgb}{0.22, 0.88, 0.08}
\definecolor{neonfuchsia}{rgb}{1.0, 0.25, 0.39}

\submitjournal{ApJ}

\shorttitle{Lithium in Red Giants}

\graphicspath{{./}{figures}}

\begin{document}

\title{Many Roads Lead to Lithium: Formation Pathways For Lithium-Rich Red Giants}

\correspondingauthor{Maryum Sayeed}
\email{maryum.sayeed@columbia.edu}

\author[0000-0001-6180-8482]{Maryum Sayeed}
\affiliation{Department of Astronomy, Columbia University, 550 West 120th Street, New York, NY, USA}

\author[0000-0001-5082-6693]{Melissa K. Ness}
\affiliation{Department of Astronomy, Columbia University, 550 West 120th Street, New York, NY, USA}
\affiliation{Center for Computational Astrophysics, Flatiron Institute, New York, NY 10010, USA}

\author[0000-0001-7516-8308]{Benjamin T. Montet}
\affiliation{School of Physics, University of New South Wales, Kensington, New South Wales, Australia}
\affiliation{UNSW Data Science Hub, University of New South Wales, Sydney, NSW 2052, Australia}

\author[0000-0002-8171-8596]{Matteo Cantiello}
\affiliation{Center for Computational Astrophysics, Flatiron Institute, New York, NY 10010, USA}

\author[0000-0003-0174-0564]{Andrew R. Casey}
\affiliation{School of Physics, University of New South Wales, Kensington, New South Wales, Australia}
\affiliation{UNSW Data Science Hub, University of New South Wales, Sydney, NSW 2052, Australia}
\affiliation{School of Physics \& Astronomy, Monash University, Clayton 3800, Victoria, Australia}

\author[0000-0002-4031-8553]{Sven Buder}
\affiliation{Research School of Astronomy and Astrophysics, Australian National University, Canberra, ACT 0200, Australia}
\affiliation{Centre of Excellence for Astrophysics in Three Dimensions (ASTRO-3D),
Australia}

\author[0000-0001-9907-7742]{Megan Bedell}
\affiliation{Center for Computational Astrophysics, Flatiron Institute, New York, NY 10010, USA}

\author[0000-0001-5228-6598]{Katelyn Breivik}
\affiliation{Center for Computational Astrophysics, Flatiron Institute, New York, NY 10010, USA}

\author[0000-0002-4670-7509]{Brian D. Metzger}
\affiliation{Department of Astronomy, Columbia University, 550 West 120th Street, New York, NY, USA}
\affiliation{Center for Computational Astrophysics, Flatiron Institute, New York, NY 10010, USA}

\author[0000-0002-3430-4163]{Sarah L. Martell}
\affiliation{School of Physics, University of New South Wales, Kensington, New South Wales, Australia}
\affiliation{Centre of Excellence for Astrophysics in Three Dimensions (ASTRO-3D),
Australia}

\author{Leah McGee-Gold}
\affiliation{Department of Physics and Astronomy, Barnard College, Columbia University, NY 10027, USA}

\begin{abstract}
Stellar models predict that lithium (Li) inside a star is destroyed during the first dredge-up phase, yet 1.2\% of red giant stars are \lirich. We aim to uncover possible origins of this population, by analysing \total Li-rich giants (A(Li) $\geq$ 1.5) in \galah\ DR3. To expose peculiar traits of \lirich stars, we construct a reference sample of \lin (\dg) stars with matched evolutionary state and fiducial iron-peak and alpha-process abundances ([Fe/H] and [Mg/Fe]). Comparing  \lirich and \dg spectra reveals systematic differences in the H-$\alpha$ and Ca-triplet line profiles associated with the velocity broadening measurement. We also find twice as many \lirich stars appear to be fast rotators (2\% with \vbr $\gtrsim 20$ \kms) compared to \dgs. On average, \lirich stars have higher abundances than their \dgs, for a subset of elements, and \lirich stars at the base of RGB have higher mean \sprocess abundances ($\geq 0.05$ dex for Ba, Y, Zr), relative to their \dgs. External mass-transfer from \textit{intermediate}$-$mass AGB companions could explain this signature. Additional companion analysis excludes binaries with mass ratios $\gtrsim$ 0.5 at $\gtrsim$ 7 AU. Finally, we confirm a prevalence of \lirich stars on the red clump that increases with lithium, which supports an evolutionary state mechanism for Li-enhancement. Multiple culprits, including binary spin-up and mass-transfer, are therefore likely mechanisms of Li-enrichment. 
\end{abstract}

\section{Introduction} \label{sec:intro}
Lithium (Li) enriched red giants remain a long-standing mystery in astrophysics \citep{WallersteinConti1969, Trimble1975, Trimble1991}. Stellar evolution predicts that atmospheric \li abundance -- determined by the interstellar medium from which the star forms -- remains fixed throughout most of the star's lifetime, but is destroyed as the star evolves from the main sequence to the red giant phase. During the first dredge-up (FDU) process \citep{iben_1968}, the outer convective layer expands and overlaps with the hotter, Li-depleted inner layers, diluting the surface \li abundance through mixing. However, \app1.2 \% red giants are found to be Li-enhanced \citep{gao_2019, casey_2019}. 

Lithium-enriched red giants were first discovered by \cite{WallersteinSneden1982}. Since then, many additional members of this population have been detected in the Milky Way field \citep[e.g.,][]{Brown1989, CharbonnelBalachandran2000, Balachandran2000, ReddyLambert2005, Gonzalez2009, Carlberg2010, CharbonnelLagarde2010, Kumar2011, Monaco2011, Ruchti2011, Kirby2012, Lebzelter2012, martell_2013, Adamow2014, daSilva2015, Dorazi2015a, Dorazi2015b, Casey2016, DelgadoMena2016, kirby_2016, Li2018, Yan2018, casey_2019, deepak_reddy_2019, gao_2019, singh_2019, Zhou2019}, clusters \citep[e.g.,][]{Sanna2020, gomez_2016b,  kirby_2016, gomez_2022a, su_2022}, and dwarf spheroidal galaxies \citep[e.g.,][]{kirby_2012}. Large-scale spectroscopic surveys have facilitated detailed studies to identify the culprit for \lirich red giants, such as \gaia-ESO \citep[e.g.,][]{Casey2016, Magrini2021}, \lamost \citep[e.g.,][]{gao_2019, wheeler2021, singh_2019, yan_2021, Ming-hao_2021, yan_2022, zhou_2022}, \galah \citep[e.g.,][]{deepak_reddy_2019, Gao2020, kumar_2020, deepak_2020, martell_2021, melinda_2021, chaname_2022}, and \rave \citep[e.g.,][]{Ruchti_2011}.

Several theories have been proposed to explain Li-enrichment, including enhancement due to asymptotic giant branch (AGB) stars \citep[e.g.,][]{cameron_1971, SackmannBoothroyd1992}, novae \citep[e.g.,][]{VigrouxArnould1979, Tajitsu_2015, Izzo2015, Starrfield1978, Molaro2016, Dearborn1989, Rukeya2017}, and cosmic ray spallation \citep[e.g.,][]{Reeves1970, OliveSchramm1992}. Others have associated \li enhancement to a specific stage in stellar evolution \citep[e.g.,][]{kumar_2020, Mallick_2023}, or other processes such as rotationally induced mixing \citep[][]{DenissenkovHerwig2004}, magnetic buoyancy \citep[e.g.,][]{Busso2007, Nordhaus2008, Guandalini2009}, wave heating \citep[e.g.,][]{Jermyn_2022}, planetary engulfment \citep[e.g.,][]{Alexander1967, SiessLivio1999a, SiessLivio1999b, VillaverLivio2009, adamow_2012}, and substellar companions \citep[e.g.,][]{King1997, Israelian2004, Israelian2009, DelgadoMena2014}.

Recent studies have associated specific formation mechanisms to certain stages of stellar evolution. \cite{casey_2019} suggested that planetary engulfment can only account for \app20\% of \lirich red giants, specifically those early on in the red giant phase. Similarly, \cite{melinda_2021} used stellar evolutionary models to show that \plen is only detectable for certain evolutionary states, while \cite{kumar_2020} suggested a universal \li production event in low-mass stars between the tip of red giant branch and the red clump.

A popular theory as to the origin of enrichment is lithium production via the Cameron-Fowler mechanism \citep{cameron_1971} where helium isotopes, $^3$He and $^4$He, must fuse together at high temperatures to produce \ber-7 ($^7$Be). However, in order to produce Li-7 ($^7$Li), $^7$Be must be transported to cooler regions to create \li via electron capture. If the surroundings are not cool enough, the \li is destroyed by proton capture to produce unstable $^8$Be, which further breaks down into two $^4$He atoms.

Motivated by the large overlap of all-sky spectroscopic and photometric surveys, we take advantage of the newly available \galah DR3 data to examine signatures of \lirich stars that might inform their formation mechanisms \citep{galah_sven_2021}. The \galah survey is especially useful as it provides measured element abundances in five nucleosynthetic families for \app$10^6$ stars near the Sun \citep{galah_survey}. The provided spectral regions include the \li and \halpha lines, where the latter is a diagnostic of activity and rotation. 

In our analysis, we use a reference sample of \lin stars to search for empirical differences in \lirich giants compared to \lin stars. We examine features directly in the spectrum itself, in particular the absorption features that are associated with chromospheric activity. We compare the prevalence of Li-enrichment as a function of evolutionary state, and examine the differences in the distributions of stellar parameters not used to construct the reference and \lirich samples. We perform a detailed analysis of the individual abundances of \lirich stars compared to \lin stars, with a focus on the \sprocess elements, to connect to the role of mass-transfer from AGB stars. We also use the individual abundances to undertake an investigation of condensation temperature trends in the \lirich sample relative to \lin stars, since condensation temperature trends have been associated with planet formation \citep[e.g.,][]{Melendez2009, Ramirez2009, Gonzalez2010}. 

Our analysis uses not only \galah but an ensemble of complementary information. This includes spectra in other wavelength regions as well as stellar parameters from \gaia and \galex, for stars common with \galah. This ensemble of information has only become recently available; its inclusion maximises the breadth of our pursuit for clues as to the formation mechanisms for Li-enrichment in giants. By combining newly available data and complementary surveys, we see evidence for multiple mechanisms for Li-enrichment. 

The paper is organized as follows. We outline the data and methods of sample construction in Section \ref{sec:obs}. We present the main findings in Section \ref{sec:results}: evolutionary state dependence in \ref{sec:evol_state}, spectral analysis in \ref{sec:spec_analysis}, evidence of stellar rotation in \ref{sec:rotation}, differences in chemical abundances in \ref{sec:abundances}, and condensation temperature trends in \ref{sec:cond_temp}. In Section \ref{sec:discussion},  we discuss the three primary results of our analysis. 

\begin{figure*}[t!]
    \centering
\includegraphics[width=0.9\textwidth]{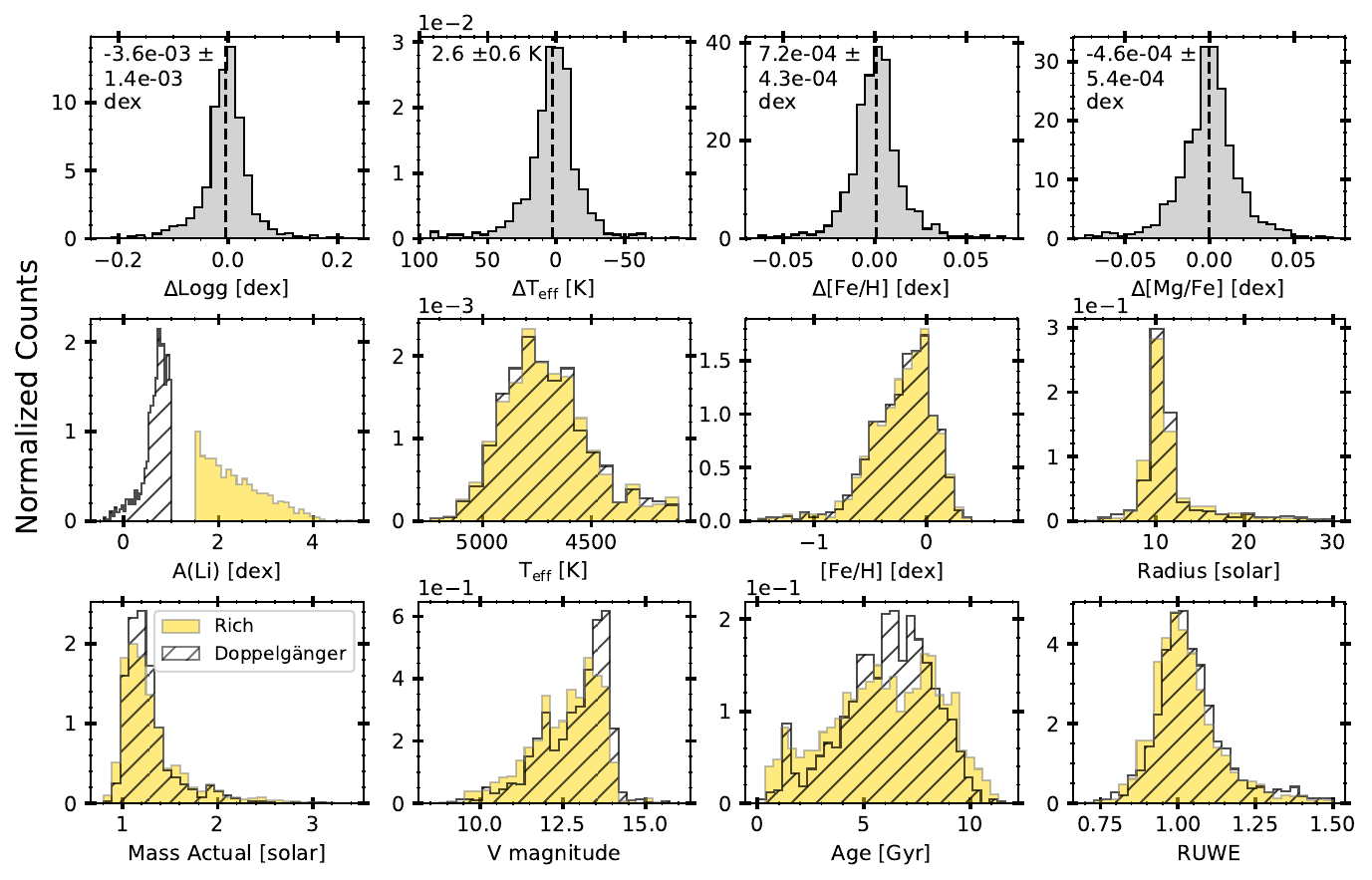}
    \caption{\textit{Top row:} 
    Normalized histograms showing the difference in stellar properties from \galah \; -- \teff, \logg, \feh, and \mgfe \; -- between \lirich stars and their \dgs. The \dgs are selected using these parameters, and have near mean zero differences and standard deviations below the uncertainties on the parameters. The text indicates the mean difference, also designated by the dashed line, and its associated error. \textit{Middle \& bottom row:} The first column in the middle row shows the distribution of lithium in the two samples. The remaining sub-panels show the normalized histograms of different parameters for the \lirich (yellow) and \dg (grey) samples in \galah. Apart from the lithium abundance, the samples are very well matched. Note that a small fraction of the \total stars falls outside of the figure shown, namely 23 stars in the radius distribution, 1 star in the $V$ magnitude distribution, and 29 stars in the RUWE distribution.}
    \label{fig:summary}
\end{figure*}

\section{Observations \& Methods}\label{sec:obs}
\subsection{Ground-Based Spectroscopy}

Galactic Archaeology with HERMES \cite[\galah;][]{galah_survey} is a high-resolution spectroscopic survey of the southern sky on the Anglo-Australian Telescope. \galah employs the High Efficiency and Resolution Multi-Element Spectrograph \cite[HERMES;][]{hermes_galah} that provides high-resolution ($R \approx 28,000$) spectra in four wavelength bands ($4713-4903$, $5648-5873$, $6478-6737$, and $7585-7887$ \ang). For our analysis, we use the latest data release \cite[\small{GALAH} DR3]{galah_sven_2021} that provides one-dimensional spectra, stellar atmospheric parameters, and up to 30 individual element abundances  for 678,423 spectra of 588,571 mostly nearby stars. 

Critically, \galah spectra covers the strong lithium absorption feature at \app6708 \ang, and \galah serves as our primary data set for investigation of \lirich stars, where a lithium abundance, [Li/Fe], was measured for 402,901 stars. Using a reference \lin sample that we construct, we look for differences that exist in \lirich stellar spectra, and derived properties, including abundances and rotation. We do this in concert with other spectroscopic and photometric survey data. We also complement the main \galah\ catalog with stellar ages, masses, and distances derived using Bayesian Stellar Parameter Estimation code (BSTEP) in \cite{sharma_bstep}.

\subsection{Sample Construction}
\label{sample_cons}
Standard nomenclature for Li-abundance, \ali \, is defined as the logarithmic abundance of \li: \ali $= \log_{10}(N_{\textrm{Li}}/N_\textrm{H}) + 12$, where $N_{\textrm{Li}}$ and $N_{\textrm{H}}$ refer to the number densities of atoms of \li and hydrogen, respectively. From measurements of [Li/Fe] and [Fe/H], we calculate the absolute lithium abundance as \ali $=$ [Li/Fe] + \feh + 1.05, where 1.05 is the solar photospheric \li abundance \citep{Asplund_2009}. Thresholds to define Li-enrichment vary \citep[e.g.,][]{gomez_2016a, gomez_2016b, deepak_reddy_2019, kumar_2020}; here, we adopt the traditional limit of \ali $\geq 1.5$ which is the mean post dredge-up value predicted by standard evolution for Population I stars. To select \lirich stars in \galah across the sky, we follow the criteria outlined in \cite{martell_2021}. We briefly summarize the steps below and direct the reader to Section 2.2 of \cite{martell_2021} for more details. Similar to their conditions, stars in our sample have:
\begin{enumerate}[i)]
    \setlength\itemsep{0em}
    \setlength\parskip{0em}
    \item a surface gravity $\log g \in [-1.0,3.2]$ dex 
    \item an effective temperature \teff $\in [3000, 5730]$ K
    \item quality flags with no known problems in the spectrum, nor in \feh measurements: \texttt{flag\_sp <= 1} \& \texttt{flag\_fe\_h = 0}
    \item \textit{\wise} W$_2$ band data quality flag of A, B, or C
    \item been excluded from LMC and SMC
    \item $E(B-V) < 0.33$
\end{enumerate}

Out of a total of 588,571 stars in \galah DR3\footnote{\url{https://www.galah-survey.org/dr3/the_catalogues/}}, 112,620 stars satisfy the above conditions. We then divide the resulting sample into \lirich and \lin groups, where `\lirich' stars have a \ali above 1.5 dex, and `\lin' (or `\dg') stars are those with upper limits for [Li/Fe] (\texttt{flag\_Li\_fe == 1}) and a \ali below 1.0 dex, which is the median Lithium abundance for the upper limit sample. This results in 1455 \lirich, and 45,163 \lin stars, where the median \ali for the latter sample is 0.67 dex. We experimented with another reference sample from which to draw \dgs from where [Li/Fe] values were measurements rather than upper limits. The different reference sample produced consistent overall results with some minor differences which we summarise in Appendix \ref{sec:app_upper_limits} and \ref{sec:app_dearth_barium}.

We then find a \dg from the available 45,163 stars for each 1455 \lirich star. We select the \dg for each \lirich star with the minimum distance in four-dimensional parameter space, where the four stellar parameters are \teff, \logg, \feh and \mgfe. Given the large set of field stars we have, this nominally selects for each \lirich star a reference object with the same evolutionary state and overall metallicity from core collapse and thermonuclear supernovae. To do this, we use the $\chi^2$ distance metric shown in Equation \ref{eq:4D} for each \lirich star, 

\begin{equation}
    \min \sum_{i=1}^{4} \frac{(x_{R,i}-x_{N,i})^2}{\sigma_{R,i}^2 + \sigma_{N,i}^2}
    \label{eq:4D}
\end{equation}
where $x_{R,i}$ is the value of the $i$th stellar parameter for the \lirich star (ie. \teff), $x_{N,i}$ is the same stellar parameter but for the \lin star, and $\sigma_{R,i}$ and $\sigma_{N,i}$ are the uncertainties on the parameters for the \lirich and \lin stars, respectively.

However, to ensure our pairs are similar in these four parameter dimensions, we also require from the \dgs that the difference in each parameter between \lirich and \lin star is less than the mean error in each parameter. The median errors are constructed using the sample of 1455 \lirich stars, which in \teff, \logg, \feh, and \mgfe are \app 100 K, 0.23 dex, 0.07 dex and 0.07 dex, respectively. This additional step ensures the \dgs are the same as their \lirich stars in the four parameters, within the typical uncertainties. Of the 1455 \lirich stars, 1158 satisfy the above conditions. However, some \lirich stars share the same \dg. We therefore remove any Li-rich star with a repeated \dg, keeping the \lirich-\dg pair with the lowest \chisq distance. Table \ref{tb:pairs} contains the \galah IDs for \total \lirich stars and their \dgs. 

Figure \ref{fig:summary} summarizes the stellar properties of the \lirich and \lin samples, showing that the \lin sample is an accurate reflection of the \lirich sample for all stellar parameters. The top row of Figure \ref{fig:summary} shows the distribution of the differences in the chosen parameters: \teff, \logg, \feh, and \mgfe. The median difference in inferred is \app9 K in \teff, 0.004 dex in \logg, 0.002 dex in \feh and 0.0005 dex in \mgfe.

Our sample includes stars in different evolutionary states. We subsequently examine our sample at the following evolutionary points: the base of the red giant branch, the red clump and the red giant branch. Prior work has suggested mechanisms of enrichment that are uniquely at the red clump stage \citep[e.g.,][]{singh_2019, deepak_reddy_2019, martell_2021, deepak_2021_jul, deepak_2021_oct}. To classify the stars in our sample as red clump or red giant members, we follow the criteria outlined in \cite{martell_2021}, and use existing \galah flags that have already categorised stars into an evolutionary state. This has been done following a Bayesian classification pipeline \citep{sharma_bstep} where, 

\begin{itemize}
    \item red clump (RC) stars have\\ \texttt{is\_redclump\_bstep} $\geq 0.50$ and $| W_2 + 1.63| \leq 0.80$
    \item red giant branch (RGB) stars have\\ \texttt{is\_redclump\_bstep} $< 0.50$ and $| W_2 + 1.63| > 0.80$
    \item base of RGB stars have \logg $\geq 2.7$ dex
\end{itemize}
Note that stars at the base of RGB are a subset of RGB stars, such that they pass the RGB selection, but they also have \logg $\geq 2.7$ dex. Implementing these conditions results in \nrgb RGB stars, \nrc RC stars, and \nbase base of RGB stars, where the mean \ali is $2.1 \pm 0.5$ dex, $2.4 \pm 0.7$ dex, and $2.1 \pm 0.5$ dex, respectively. 

\subsection{\gaia DR3 \& \galex} \label{sec:gaiadr3}
\begin{deluxetable}{cc}[t!]
\tablecaption{\galah ID for \lirich stars and their \dgs.\label{tb:pairs}}
\tablehead{\colhead{Rich} & \colhead{Doppelgänger} 
} 
\startdata
131118002901313 & 170216002301393 \\
131120002001376 & 150103003501218 \\
131123003501064 & 150831002501029 \\
131218002401174 & 161108002101211 \\
140112002301046 & 161228003001222 \\
140209002201006 & 170508004801045 \\
140209002202072 & 160417002201138 \\
140303000402167 & 151111002101361 \\
140307003101263 & 160327005101035 \\
140309003101259 & 181222001801311 \\
... & ... \\
\enddata
\end{deluxetable}

We also use \gaia data for our sample of \galah stars to analyze regions of the stellar spectra not covered by \galah. \gaia Data Release 3 (DR3) provides near-infrared (845–872 nm) spectra at medium-resolution ($R \approx 11,500$), measured with the Radial Velocity Spectrometer (RVS) for 999,645 sources and centered on the Ca-triplet \citep{gaia_collab, Cropper2018, gaiadr3}. The Ca-triplet core is particularly interesting as a marker of magnetic activity in the upper layers of the chromosphere \citep{Lanzafame_2022}. We therefore wish to compare this feature in \lirich and \lin stars. We find 673 stars in our \lirich sample with available RVS spectra. Note that while it is possible to model magnetic field strength from variation in the Ca-triplet lines \citep[e.g.,][]{Carlin2015}, this is beyond the scope of this work.

We find \dgs for these stars in \gaia spectra in order to compare the Ca-triplet feature in the two samples; however, we now use the spectra to find \dgs, rather than the stellar parameters. While we can not explicitly select only \lin stars in our reference \dg set, as there is no lithium measurement from \gaia, the candidate \dgs are most likely to be \lin, given that only \app1\% of giants are \lirich. The criteria and steps taken to create the \dg sample are described below:
\begin{deluxetable}{cc}[t!]
\tablecaption{\gaia DR3 ID for \totalrvs \lirich stars and their \dgs with \gaia RVS spectra.\label{tb:gaia_rvs}}
\tablehead{\colhead{Rich} & \colhead{Doppelg\"anger}} 
\startdata
1753307668590290304 & 4986973393100811776 \\
2535920013509617152 & 4844263690123001856 \\
2536027971807637248 & 3575326288398736256 \\
2549549422208614912 & 5223284589673766272 \\
2549897589437503488 & 6407598113323451648 \\
2597169751843870336 & 5771702435348541184 \\
2608309007224057856 & 6848264300515911552 \\
2613265021526634624 & 6541769524395228032 \\
2622394708952964096 & 6913587244096921344 \\
2679552786563683328 & 5476435420998964480 \\
... & ... \\
\enddata
\end{deluxetable}

\begin{enumerate}[i)]
    \setlength\itemsep{0em}
    \setlength\parskip{0em}
    \item We remove the 673 \lirich from the RVS catalog; 998,972 stars remain.
   \item To reduce the search space for \dgs for each \lirich object, we generate a list of possible \dgs for each \lirich star with differences in \teff and \feh between the \lirich and \dg  within the mean uncertainties ($<$ 100 K for \teff and $<$ 0.15 dex for \feh). We also ensure the signal-to-noise (SNR) of the \dg is above 50; this results in a list of 182,898 possible \dgs for all 673 \lirich stars. 
   \item We use a $\chi^2$ distance metric on the spectra but exclude the Ca-triplet region. We compare each \lirich star to the subset of candidate \dg stars using this distance metric, arriving at a $\chi^2$ distance for each possible \dg for each \lirich star. The numerator terms are the normalised flux of the pair of stars and the denominator terms are the corresponding flux uncertainties. The wavelength regions we calculate this over are $8477-8492$ \ang, $8507-8532$ \ang, and $8600-8653$ \ang. For each \lirich star, we find a set of \dg stars with a \chisq value below 2 times the degrees of freedom (ie. $2\times N_\lambda$).
   \item For all \dg stars that satisfy the above condition, we select a random star from the shortlist as the \dg, resulting in a final \totalrvs \lirich stars with \gaia RVS spectra matched with a \dg.
\end{enumerate}
Table \ref{tb:gaia_rvs} contains the \gaia DR3 IDs for the \totalrvs \lirich stars and their \dgs with \gaia RVS spectra.

Furthermore, to search for any \lirich objects with significant signatures of stellar activity, we also look for emission in the ultraviolet (UV). We use NASA's Galaxy Evolution Explorer (\galex), and its associated All-Sky Imaging Survey (AIS), to examine the UV emission in our stars \citep[]{galex_martin, galex_ais}. Within \galex\footnote{using MAST to cross-match our sample with GR6+7 from all catalogs (AIS, MIS, DIS)}, we find \richgalex of our \lirich and \poorgalex \lin stars using 2\arcsec \; for the cross-match. Object IDs from the three surveys mentioned -- \galah, \gaia and \galex \;-- are included in Tables \ref{tb:full_rich} and \ref{tb:full_poor} for the \lirich and \dg samples.

\begin{figure}[t!]
    \centering
    \includegraphics[width=0.5\textwidth]{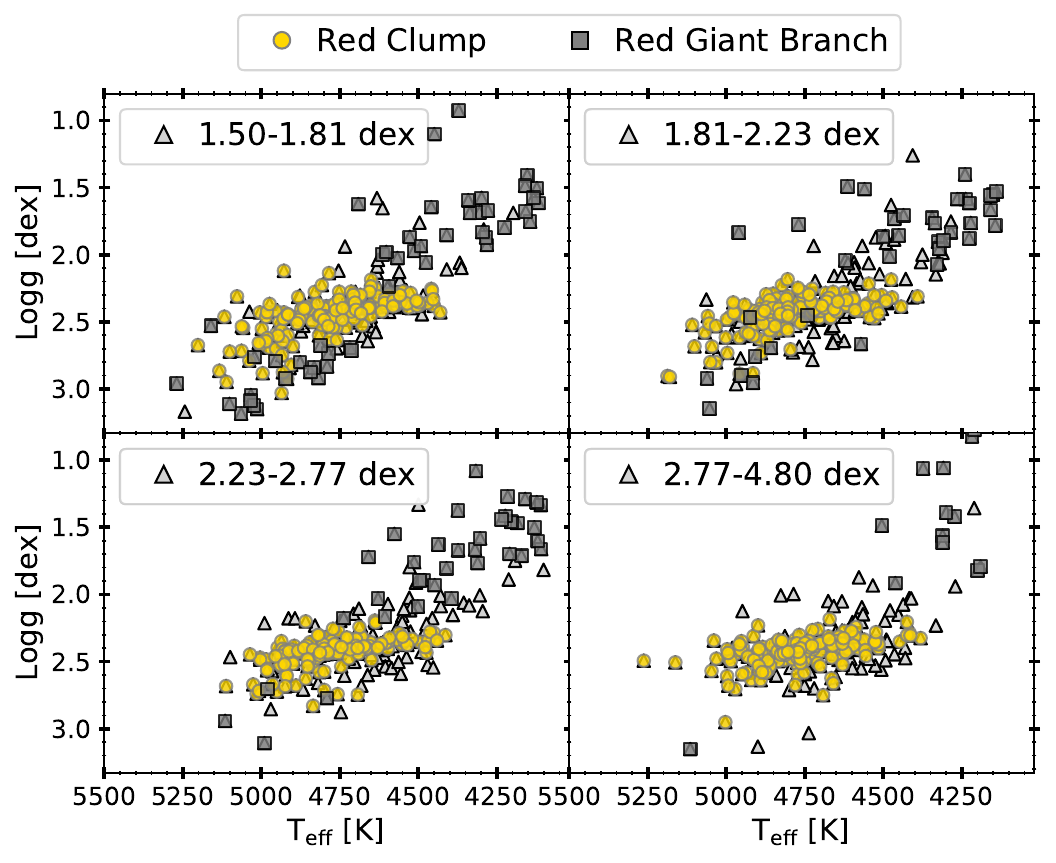}
    \caption{Kiel diagram of \lirich stars divided into four equal bins of \li abundance where the \ali of data points in each subplot is indicated in the legend. The grey squares and yellow circles show stars on the red giant branch and red clump, respectively, based on criteria from \cite{martell_2021}. The distribution of \logg of stars in our sample becomes more centered around \logg of \app2.4 dex as stars become more \lirich, indicating that super$-$\lirich stars are more likely to be red clump stars. Also see Figure \ref{fig:random_logg}.}
    \label{fig:teff_vs_logg}
\end{figure}

\begin{figure}[t!]
    \centering
    \includegraphics[width=0.5\textwidth]{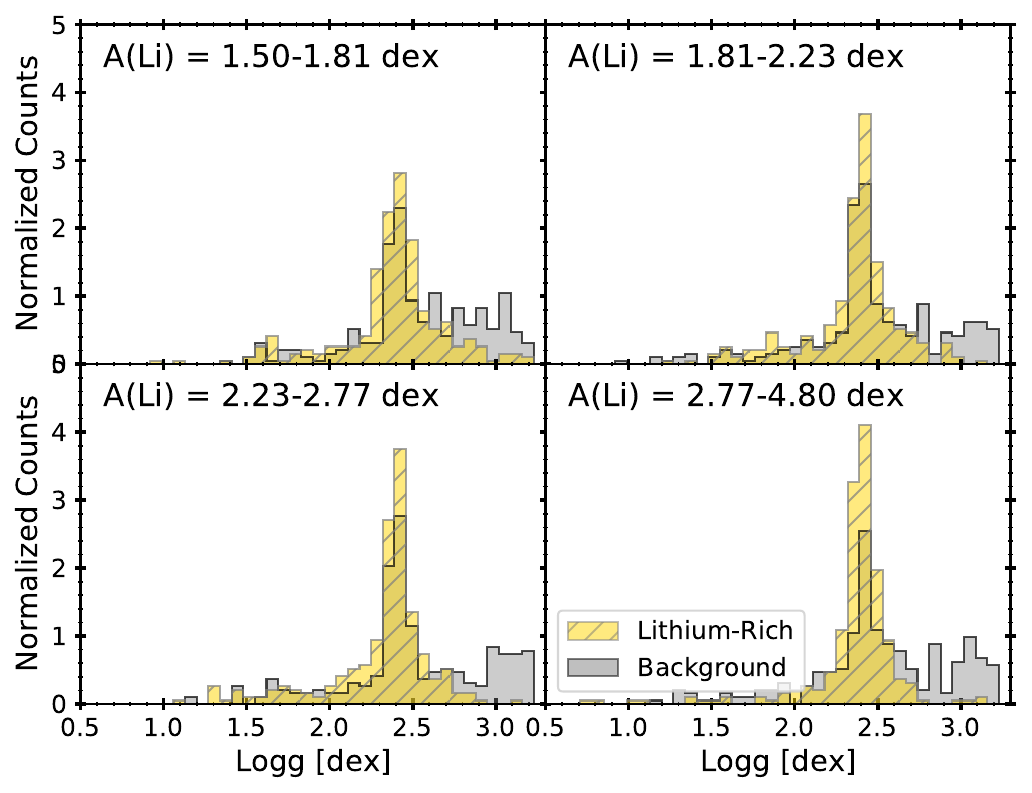}
    \caption{Histograms showing randomly sampled \logg in \galah in four bins of \li abundance to demonstrate that the background \galah sample is not preferentially centered at a specific \logg, and that our finding that red clump stars are more likely to be \lirich (shown in Figure \ref{fig:teff_vs_logg}) is real.}
    \label{fig:random_logg} 
\end{figure}

\section{Results} \label{sec:results}

We summarize our findings in five main results: 
\begin{enumerate}[i)]
    \setlength\itemsep{0em}
    \setlength\parskip{0em}
    \item preferential red clump membership as a function of Li-enrichment,
    \item evidence of net differences in the H-$\alpha$ and Ca-triplet line profiles, 
    \item differences in the mean rotation of a subset of \lirich giants relative to their \dgs,
    \item difference in mean element abundances for some elements for the \lirich population, in particular signatures in neutron-capture elements and at the base of the red giant branch, and
    \item an absence of any difference between the \lirich and \lin stars' trends in condensation temperature ranked element abundance values.
\end{enumerate}

\begin{figure*}[t!] 
    \centering
    \includegraphics[width=0.95\textwidth]{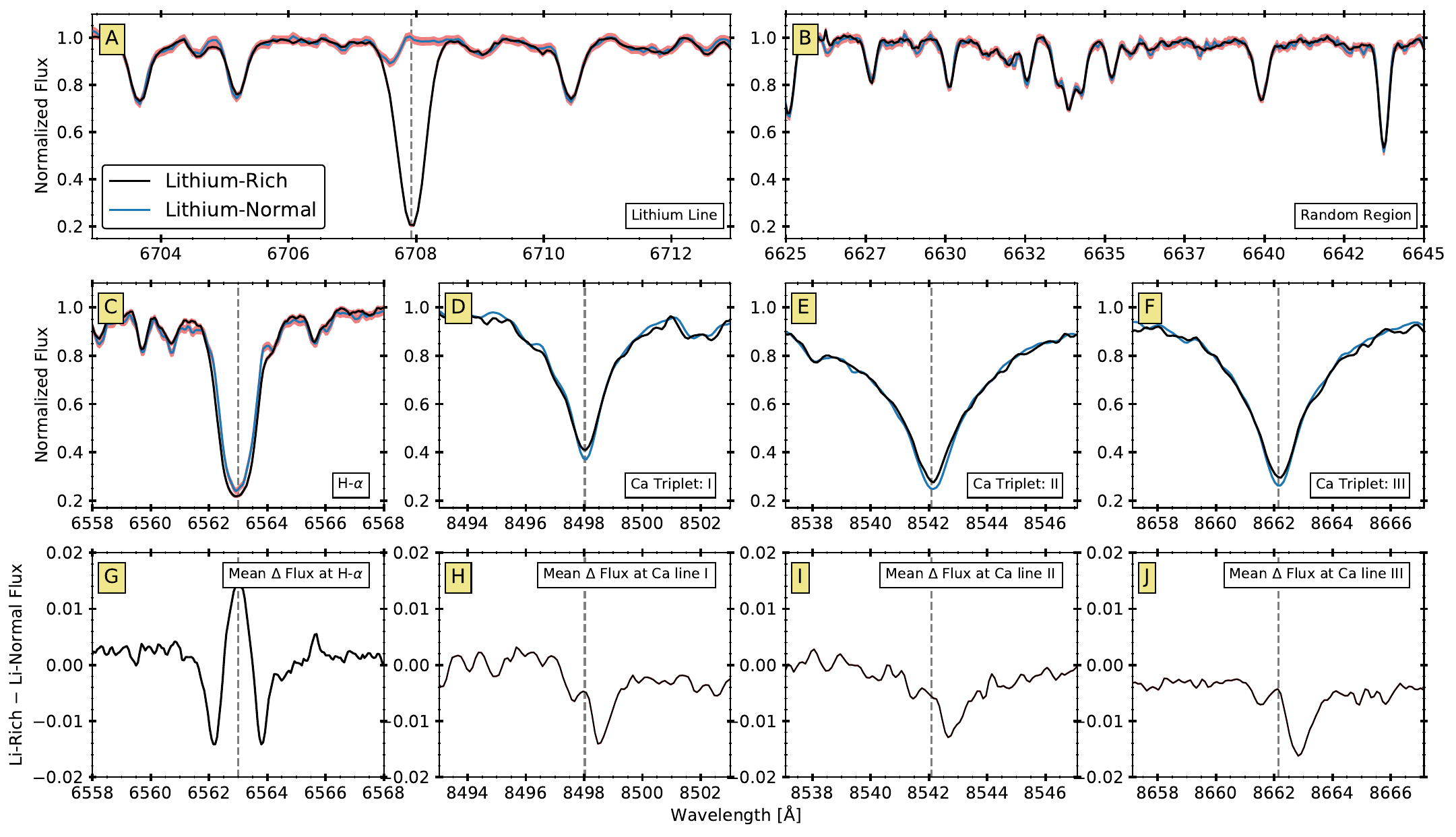}
    \caption{Examples of spectra from \galah and \gaia of \lirich and \dg pairs. Plots A, B, C and G show \galah spectra, and plots D-F and H-J show \gaia RVS spectra. The black and blue curves show data for \lirich and \lin stars, respectively. The red shaded region represents the reported uncertainty on the flux.
    The chosen \lirich star is \galah ID: 160919001601113 (or \gaia DR3 ID 6731814646069597312). Its \dg in \galah is \galah ID 160530001601265 (or \gaia DR3 ID 5242820609200120576), and its \dg in \gaia RVS data (panels D-F and H-J) is \gaia DR3 ID 6345552672167837312.
    \textbf{Panel A:} region of \galah spectra centered around the Li-line at 6707.926 \AA; the \lirich star clearly has a strong lithium absorption feature not present in the \lin star.
    \textbf{Panel B:} arbitrary region of \galah spectra to show the spectra look nearly identical.
    \textbf{Panel C:} \galah spectra centered around the \halpha line at 6562.79 \AA. 
    \textbf{Panel D, E, F:} \gaia spectra centered on the Ca-triplet lines.
    \textbf{Panel G:} average difference in flux for all \total \lirich and \lin stars in \galah.
    \textbf{Panel H, I, J:} average difference in flux at the Ca-triplet lines between all \totalrvs \lirich and \lin stars in \gaia with available \gaia RVS data. The error in flux is plotted under each curve.} 
    \label{fig:eg_spectrum}
\end{figure*}
\subsection{Evolutionary State Analysis}
\label{sec:evol_state}

To investigate the correlation between evolutionary state and Li-enhancement, we divide our sample of \lirich stars into bins of lithium with equivalent numbers of stars within each bin, shown on a a Kiel diagram in Figure \ref{fig:teff_vs_logg}. We see that the most \lirich stars (ie. \ali $= 2.8-4.2$) are preferentially red clump stars. In fact, we see overall that as the \ali increases, a larger relative fraction of the stars fall on the clump. More specifically, the fraction of RGB stars decreases from 18\% to 5\%, while the fraction of RC stars increases from 56\% to 67\%, for higher \li abundances. 

Moreover, in Figure \ref{fig:random_logg} we show that this trend is not due to the survey selection function. To demonstrate this, we select \galah stars with a \teff and \logg range of our \lirich sample, and randomly sample \total `background' stars in \galah from a sample of 112,620 total stars, repeated 100 times. As seen in Figure \ref{fig:random_logg}, we see that the background sample is not preferentially centered at a given \logg as a function of \li abundance, contrary to our \lirich sample. Therefore, our finding of increasing numbers of red clump stars at higher \li abundance in Figure \ref{fig:teff_vs_logg} is not a consequence of \galah's selection function. 

\subsection{Spectral Analysis}
\label{sec:spec_analysis}

\begin{figure}[t!]
    \centering
    \includegraphics[width=0.5\textwidth]{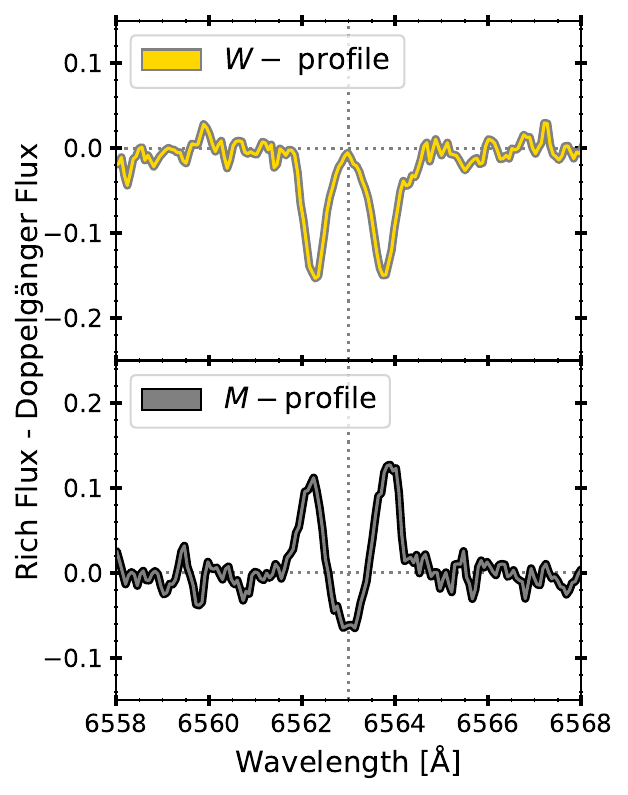}
    \caption{
    Difference in spectral flux between \lirich and \dg star centered on the \halpha line. 
    \textit{Top:} difference between a \lirich star and its \dg where the resulting spectrum creates a $W-$profile, corresponding to a \lirich star with emission in the core relative to its \dg, and with deeper absorption on either side.
    \textit{Bottom:} difference between a \lirich star and its \dg where the resulting spectrum creates a $M-$profile, corresponding to a \lirich star with emission in the wings relative to its \dg, and with deeper absorption in the core.}
    \label{fig:w_vs_m}
\end{figure}

\begin{figure}[t!]
    \centering
    \includegraphics[width=0.5\textwidth]{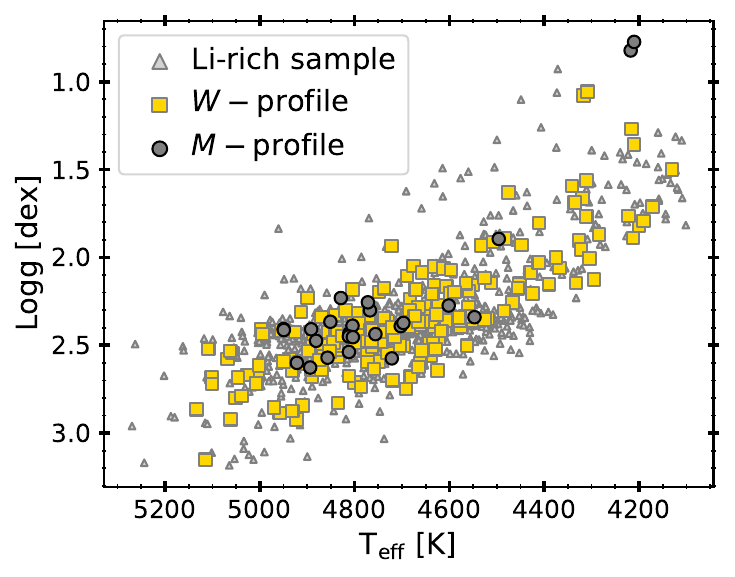}
    \caption{Stars classified as $W-$ or $M-$ in the \lirich sample on a Kiel diagram.}
    \label{fig:hr_mw}
\end{figure}

\begin{figure*}[t!]
    \centering
    \includegraphics[width=0.99\textwidth]{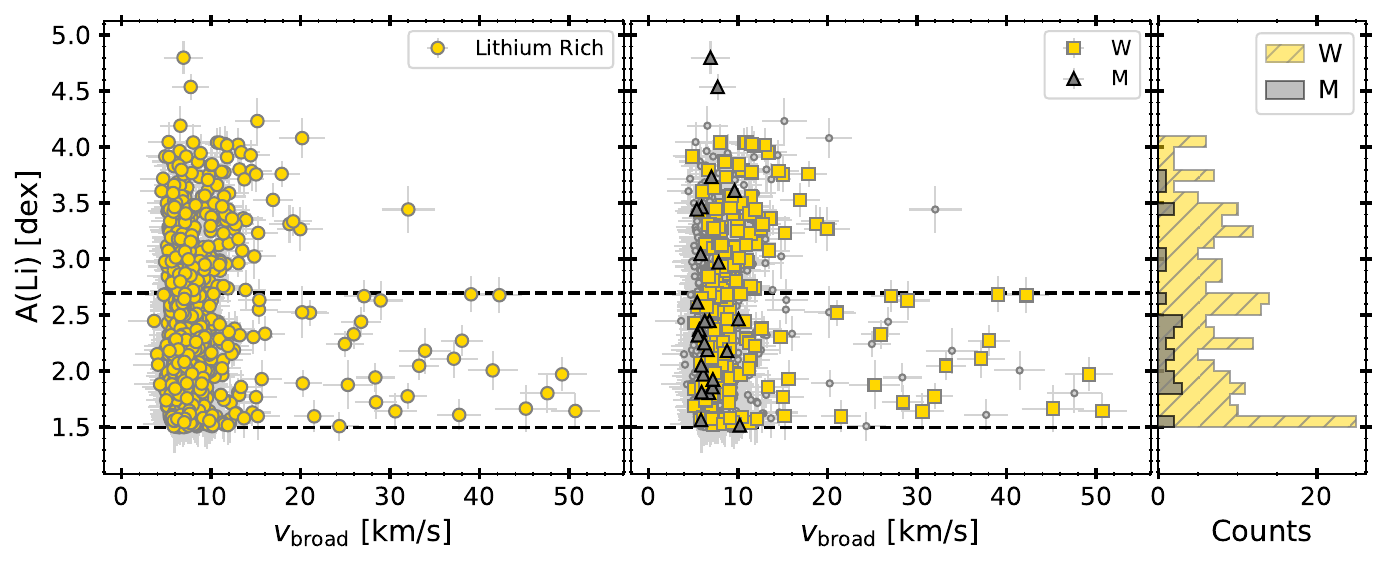}
    \caption{Distribution of \li abundance, \ali, as a function of \vbr (in \kms). \textit{Left:} data points show the \lirich sample. \textit{Middle:} same sample as left panel, separated into stars that create $W-$profiles (yellow squares) or $M-$profiles (grey triangles) in their \halpha \dg difference. The dashed lines at \ali$= 1.5$ dex and \ali$= 2.7$ dex are shown for reference. \textit{Right:} distribution of the two profiles -- $W$ and $M$ -- as a function of \ali. Stars with $W-$profiles have higher \vbr as compared to stars with $M-$profiles. This makes physical sense as deeper wing profiles correspond to faster rotation.}
    \label{fig:vbroad_sarah}
\end{figure*}

The \galah and \gaia spectra have features -- notably the \halpha and Ca H \& K lines, respectively -- that can probe emission and chromospheric activity, and reveal stellar rotation due to their broad and strong line profiles that are formed in the chromosphere \citep[e.g.,][]{isik2023}.  We examine these features in the \lirich and \lin samples. In Figure \ref{fig:eg_spectrum}, we show examples of spectra centered around the \li and \halpha lines in \galah, and the Ca-triplet in \gaia. In Panels G-J, we show the difference in flux between \lirich and \lin stars averaged over the complete sample (ie. \total in \galah and \totalrvs in \gaia), thereby reducing the sampling noise at each wavelength to (\app 0.0003 in \galah and \app 0.002 in \gaia spectra). We see statistically significant differences at the \halpha and Ca-triplet lines, and an asymmetry in the latter which could be caused by surface velocity structure \citep[e.g.,][]{gray_1980, mallik_1997, Nieminen_2017}. These profiles in Panels G-J are potentially consistent with systematic differences in activity in the stellar chromosphere \citep[e.g.,][]{sneden2022}, differences in stellar rotation, or binary fractions between the \lirich and reference samples; all would manifest in similar profiles seen in the bottom panel of Figure \ref{fig:eg_spectrum}. 

If the Cameron-Fowler mechanism is responsible for Li-enriched stars, this should also correspondingly enhance another product of the reaction, namely beryllium \citep{cameron_1971}. By pooling our spectra, we attempted to detect any small differences in absorption strength at the \ber line\footnote{From NIST: \url{https://physics.nist.gov/PhysRefData/Handbook/Tables/berylliumtable2.htm}} at 4828.159 \ang between the \lirich and \lin stars. Despite stacking spectra of \total \lirich stars to increase the signal, we made no detection of beryllium. However, there have been detections of \ber in stars, but at a line not within the \galah spectral range \citep[e.g.,][]{Gilmore_1991, Boesgaard_2009, Giribaldi_2022, Smiljanic_2022}. 

\subsubsection{\halpha Line Profiles} \label{sec:halpha}

Panel G of Figure \ref{fig:eg_spectrum} shows the mean difference in the core of the \halpha line in the \lirich stars as compared to their \dgs. To investigate this finding in more detail, we examine this \halpha feature as a function of \feh and \ali for both \lirich and \lin stars, but find no significant dependence of \feh and \ali on the \halpha feature when binning in \feh and \ali. 

However, by examining the difference in the spectra for individual pairs in more detail, we noticed differing shapes of the spectrum at the \halpha line. We subsequently categorized each pair difference into a `$W-$profile', `$M-$profile', or neither; Figure \ref{fig:w_vs_m} shows an example of a $W-$profile and $M-$profile. Of our \total \lirich stars, 20\% of the sample (\wstars stars) produced a $W-$profile with their \dg, 2\% of the sample (\mstars stars) produced an $M-$profile, and 78\% were unclassified. Figure \ref{fig:hr_mw} shows the distribution of $W$ and $M$ designations on a Kiel diagram. Interestingly, almost all the $M-$designations fall on the red clump, at \logg $\sim 2.4$ dex, while the $W-$designations span the entire giant branch. 

As seen in Figure \ref{fig:w_vs_m}, $W-$designations correspond to deeper flux in the wings in the \lirich stars and higher flux in the core, while $M-$designations refer to deeper flux in the core and higher flux in the wings for the \lirich star. Based on spectral shapes, these differences could be explained by respectively faster rotation of the \lirich star compared to the \dg ($W-$profile), and slower rotation of the \lirich star compared to the \dg ($M-$profile). The $W$ and $M$ profiles would also be produced if one of the pairs is in a (barely-detectable) spectroscopic binary, the $W-$profile when the \lirich star is a binary and the $M-$profile when the \dg is a binary; we investigate this further in Section \ref{sec:flag32}. Interestingly, these profiles show asymmetry around the center of the line, with the higher wavelength wing profile of the $M-$designations showing a larger difference than the lower wavelength wing.

Similarly, we inspect differences in the \gaia RVS spectra between the \lirich and \lin stars around the Ca-triplet lines. We similarly classify the differences into $W$ or $M$ through visual inspection. Of the 667 stars with \gaia spectra, 53 of them are classified as $W$ and 34 as $M$ relative to the \dg. 

These differences in the line profiles in both the \halpha and Ca-triplet between the \lirich stars and their \dgs could be  rotation driven (e.g., shallower and broader lines), and we directly investigate the measured rotation in the next section. Under this interpretation, the higher fraction of $W-$profiles (\app 10:1) is consistent with a fraction of \lirich stars having a net faster rotation than their \lin counterparts in the overall population. Conversely, binarity could masquerade as a measured faster rotation for a single star. However, further follow up work with multi epoch spectra is needed to differentiate between these scenarios.

\subsection{Stellar Rotation Rate} \label{sec:rotation}

\subsubsection{\vbr} \label{sec:vbroad}
Since the spectra indicate the possibility of some rotation differences between the \lirich and \lin samples, we examine the rotation of the \lirich and \lin samples via the broadening velocity (\vbr) measurement for each star. \galah provides \vbr measurements which encompasses macroturbulence and rotational velocities (fitted with \vsini). This indicates a subtle signature of a fraction of faster rotators in the \lirich population. Note that we have confirmed that the rotational broadening estimates, which could serve as a parameter to model the empirical resolution of the data, show no systematic difference with observing date, and between \lirich and \lin samples.

\begin{figure}[t!]
    \centering
    \includegraphics[width=0.5\textwidth]{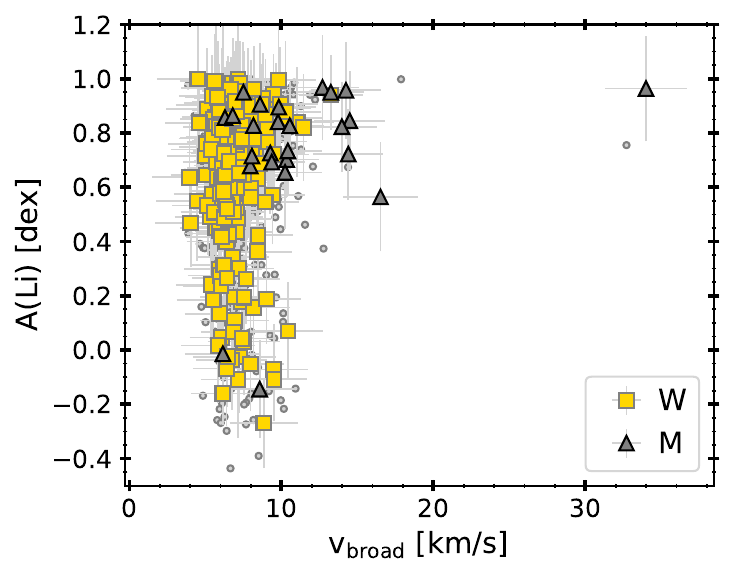}
    \caption{Distribution of \li abundance, \ali, as a function of \vbr (in \kms) for \dgs, separated into \lirich stars that show $W-$profiles (yellow squares) or $M-$profiles (grey triangles) in their \halpha \dg difference. \lin stars where their \lirich counterpart has a $M-$profile have high \vbr than \dgs where the \lirich star has a $W-$profile, opposite of the trends seen in the \lirich population in Figure \ref{fig:vbroad_sarah}.}
    \label{fig:vbroad_dg}
\end{figure}

Figure \ref{fig:vbroad_sarah} shows the distribution of \lirich stars as a function of \vbr for the complete sample, $W$ and $M$ profiles, and Figure \ref{fig:vbroad_dg} shows corresponding distribution for the \dgs (similar to the left panel of Figure \ref{fig:vbroad_sarah}). From Figures \ref{fig:vbroad_sarah} and \ref{fig:vbroad_dg}, the vast majority of stars in both the \lirich and \dg populations, have a measured \vbr \app $5-15$ \kms (with typical \vbr measurement uncertainties of \app2 \kms). However, we see a very small fraction of anomalously high rotators. In fact, we find ten times as many fast rotators in the \lirich sample compared to the \dg population; in the \lirich sample, \app2.6\% of the stars have \vbr $\geq 20$ \kms  (29/\total), compared to only 0.2\% in the \lin sample (2/\totaldg). Following typical classifications, a \vsini $\gtrsim 10$ \kms is commonly defined as a fast rotator on the RGB \citep[e.g.,][]{Carney2008, Patton_2023}, which is a generous threshold for fast rotation, but also robustly accounts for measurement uncertainty in the reported \vbr. 

From the middle and right panels of Figure \ref{fig:vbroad_sarah}, we see a clear difference in the \vbr distributions of the $W$ and $M$ designations. $W-$designations show a larger relative fraction of stars with lower lithium enhancement, which are preferentially the stars rotating faster than their \dgs. Conversely, $M-$designations all fall to the lower boundary of the \vbr distribution; these are the stars for which the \lirich star is rotating slower than its \dg. The majority of the fast rotators (\vbr $\geq$ 20 \kms) are classified using the \halpha feature; of the stars with \vbr $\geq 20$ \kms, 59\% (17/29) are $W-$designations, and the rest are neither, which further suggests that $W-$designations are more likely to be \lirich. In Figure \ref{fig:vbroad_dg}, the anomalously high rotators are the reverse designation, $M$, as in this case typically the \dg is the faster rotator of the pair. 

\begin{figure}[t!]
    \centering
    \includegraphics[width=0.5\textwidth]{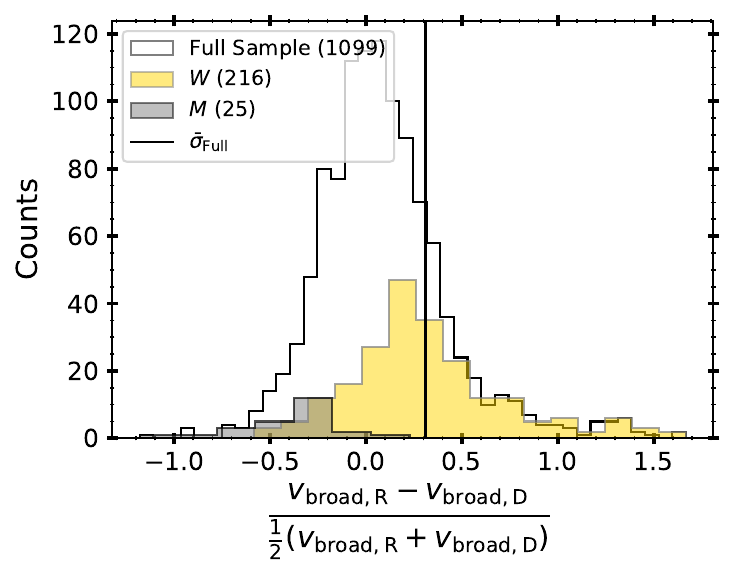}
    \caption{Fractional difference in broadening velocity (\vbr) between a \lirich star and its \dg for all stars in our sample (solid black bars), stars with $W-$profiles (yellow bars), and stars with $M-$profiles (grey bars) in their \dg difference. The solid line shows the 1$-\sigma$ standard deviation expected around a zero mean based on the measurement uncertainty alone for the full sample.}
    \label{fig:vbroad_diff}
\end{figure}

In Figure \ref{fig:vbroad_diff}, we show the fractional difference in \vbr between the pairs of stars (\lirich $-$ \dg) for the full sample, and for the two designations, $W$ and $M$. The median fractional difference in \vbr is $0.05 \pm 0.01$, $0.25 \pm 0.03$, and $-0.37 \pm 0.03$ for the full \lirich sample, $W-$designations and $M-$designations, respectively. The net rotation of the \lirich stars is therefore marginally higher than the \dg sample, overall. 

Figure \ref{fig:vbroad_diff} clearly shows that \lirich stars with $W-$profiles have higher \vbr relative to their \dg, while \lirich stars with $M-$profiles have a lower \vbr than their \dg.  That is, \lirich stars that show deeper \halpha wings than their \dgs have correspondingly higher rotation, which makes sense as rotation deepens the wings of the \halpha line \citep[e.g.,][]{Petrenz1996}. Conversely, $M-$designations, which have shallower wings and deeper cores in the \lirich stars compared to their \dgs, show lesser broadening. Note that there are almost ten times as many $W-$designations than $M-$designations, which is also expressed as a skew seen in the distribution of the full sample seen Figure \ref{fig:vbroad_diff}.

\subsubsection{UV and IR and Gaia Photometry} \label{sec:uv}
\begin{figure}[t!]
    \centering
    \includegraphics[width=0.5\textwidth]{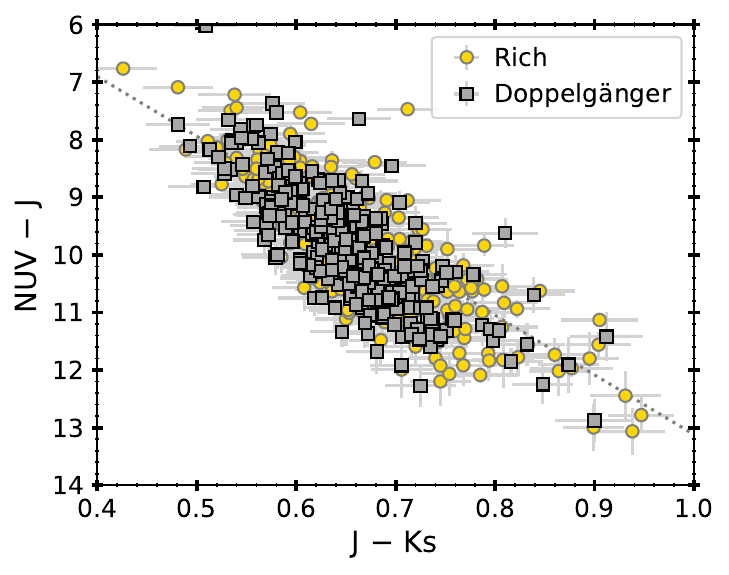}
    \caption{Colour-colour diagram of our \lirich (yellow circles) and \lin (grey squares) samples, where the dotted line shows the reference NUV excess. The close overlap in the distribution of the \lirich and \lin samples suggests no UV excess for \lirich stars detected by \galex.}
    \label{fig:uv}
\end{figure}

Given the well-known correlation between rotation and stellar activity \citep[e.g.,][]{Wilson1966, Kraft1967, Noyes1984, Soderblom1993}, we look for any relationship between ultraviolet (UV) emission and \li abundance.
\cite{dixon_2020} derived an empirical relationship between near-UV (NUV) excess and rotational velocity (\vsini) for their sample of 133 stars in \apogee and \galex \citep{apogee_dr16}; we use a similar method to look for NUV excess in our \lirich sample. In Figure \ref{fig:uv}, we show our \lirich and \dg sample on a colour-colour diagram. The dotted black line shows the reference UV excess activity \citep{FindeisenHillenbrand2010, dixon_2020} described by the following:
\begin{equation}
    NUV - J = (10.36 \pm 0.07)(J - K_s) + (2.76 \pm 0.04)
    \label{eq:uv_fit}
\end{equation}
We use $J$ and $K_s$ measurements from 2MASS, and NUV measurements from \galex DR6. As seen in Figure \ref{fig:uv}, we see no difference in the distribution of points between our \lirich and \lin samples. 

We also looked for differences in infrared (IR) excess between the \lirich and \lin samples using W1 and W4 magnitudes from \wise \citep[][]{wise_catalog, rebull_2015}, but found no differences in distribution between \lirich and \lin samples, or between $W$ and $M$ designations. We also checked for indications of photometric variability using \gaia DR3 photometry, specifically in the $G$, $G_\textrm{BP}$ ($400-500$ nm) and $G_\textrm{RP}$ ($600-750$ nm) bandpasses. We found no difference in distributions between the \lirich and \dg samples. Finally, we searched for correlations between lithium abundance and the re-normalized unit weight-error (RUWE). RUWE is the magnitude and colour-independent re-normalization of the astrometric \chisq fit in \gaia DR2, which is sensitive to close binaries \citep[e.g.,][]{evans_2018, gaia_2018, Lindegren_2018, berger_2020}. In doing a best fit between \vbr and RUWE, we found no indication that \vbr, nor \ali, are correlated with RUWE given a negligible slope between \ali vs. \vbr, and \ali vs. RUWE. 

\subsection{Abundance Analysis}
\label{sec:abundances}

\begin{figure*}[t!]
    \centering
    \includegraphics[width=0.9\textwidth]{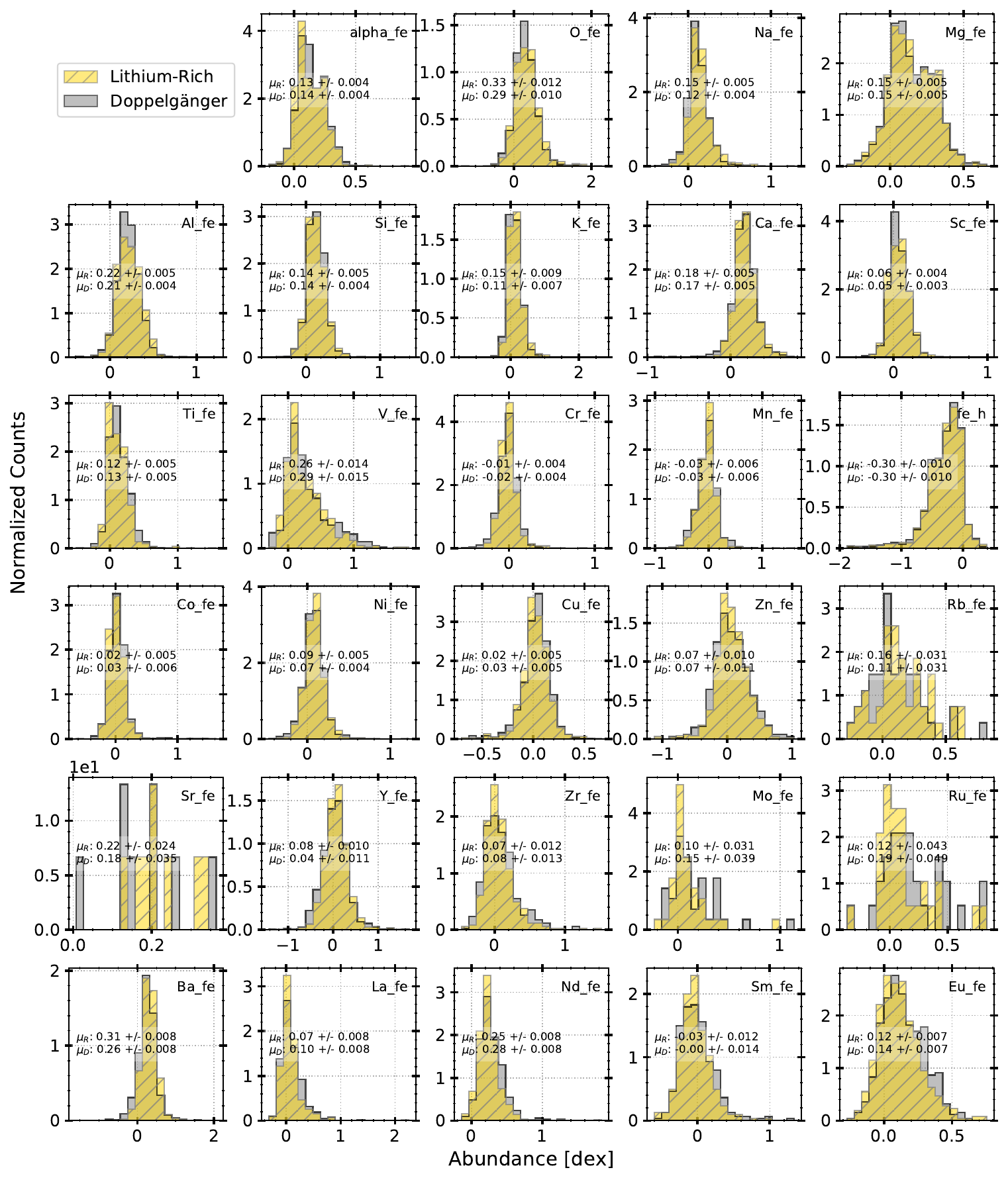}
    \caption{Distribution of 29 abundances for the \lirich (yellow hatched) and \lin (grey) sample. The mean abundance and confidence on the mean for the \lirich and \dg samples are indicated in text with $\mu_R$ and $\mu_D$, respectively.}
    \label{fig:delta_abundance}
\end{figure*}

The multiple individual abundances measured for \galah\ enables us to compare the distribution of abundances between \lirich and \dg samples. We examine the following elements: Al, $\alpha$, Ba, Ca, Co, Cr, Cu, Eu, Fe, K, La, Li, Mg, Mn, Mo, Na, Nd, Ni, O, Rb, Ru, Sc, Si, Sm, Sr, Ti, V, Y, Zn, and Zr. These elements can be divided into the following five categories \citep[][]{Buder_thesis}:
\begin{enumerate}[i)]
    \setlength\itemsep{0em}
    \setlength\parskip{0em}
    \item $\alpha$ elements -- Ca, Mg, Si, Ti: created hydrostatically and explosively (depending on the element), by $\alpha$-particle capture in massive stars, and released into the ISM by core collapse supernovae (SNe II)
    \item odd$-Z$ elements -- Al, K, Na: produced during supernovae type Ia (SNe Ia) and SNe II, and explosive C, O, and Ne burning
    \item iron-peak elements -- Fe, Co, Cr, Cu, Mn, Ni, Sc, V, Zn: produced during supernovae
    \item \sprocess elements -- Ba, La, Rb, Sr, Y, Zr: produced by neutron-capture and decay processes, most likely in AGB stars or massive star winds
    \item $r-$process elements -- Ce, Eu, Nd, Ru: produced by neutron-capture and decay processes, hypothesized to form during kilonovae, magnetorotational supernovae and collapsars \textbf{\citep[e.g.,][]{Kobayashi2020}}
\end{enumerate}

\begin{figure*}[t!]
    \centering
    \includegraphics[width=0.9\textwidth]{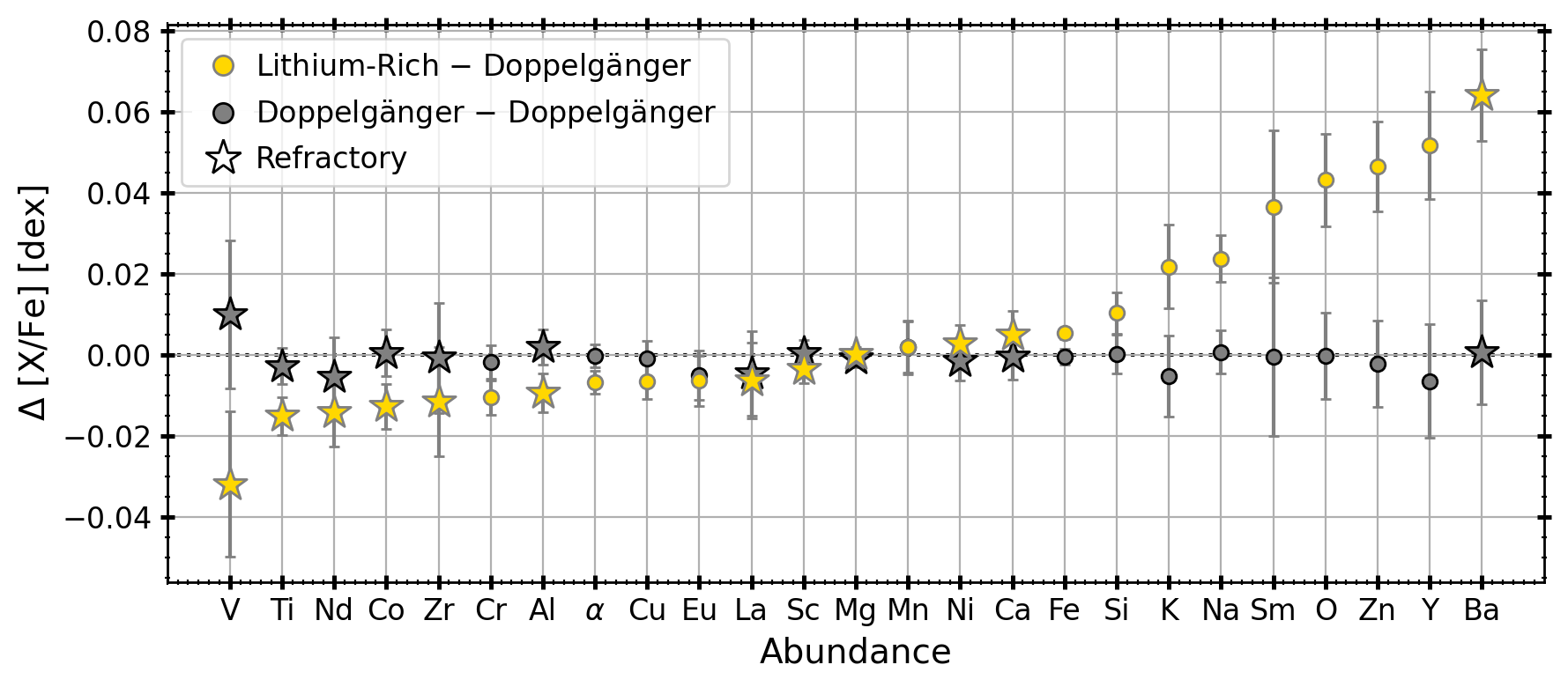}
    \caption{Mean difference in abundance between the \lirich and \dg sample as a function of elements. The star markers indicate the refractory elements, and error bars are shown in grey. The yellow points show the difference between a \lirich and \lin star, and the grey points show the difference between two random \dgs of the \lirich star. Since each point represents the average of 10 samples, the differences between \lirich and \lin sample are more significant compared to differences between the \dg-\dg sample.}
    \label{fig:diff_in_abundance}
\end{figure*}

For this analysis, we only use stars with good quality flags for a given element (\texttt{flag = 0}), which reduces the \lirich sample of \total stars by less than $15\%$ for most elements, $15-50\%$ for Eu, La, Nd, V, Zn, and Zr, by \app$70\%$ for Sm, and by more than $90\%$ for Mo, Rb, Ru, and Sr. Figure \ref{fig:delta_abundance} shows histograms of the \lirich and \lin populations for each element, with the mean and confidence on the mean ($\sigma/\sqrt{N}$) also reported in each sub-panel. By inspection of the histograms and summary statistics, namely the mean and 1$-\sigma$ standard deviation, the \lirich and \lin samples are near-identical for many elements. In fact, \app20\% of elements have the same mean and 1$-\sigma$ standard deviation within uncertainties, for the \lirich and \lin populations. Within the $2-\sigma$ uncertainty on the mean, \app50\% of elements\footnote{[$\alpha$/Fe], [Al/Fe], [Ca/Fe], [Co/Fe], [Cu/Fe], [Fe/H], [Mg/Fe], [Mn/Fe], [Mo/Fe], [Rb/Fe], [Ru/Fe], [Sm/Fe], [Sr/Fe], [V/Fe]} have the same mean values in the \lirich and \lin samples. It is perhaps not surprising that most of the supernovae Ia and II element distributions are the same within uncertainties, as \feh and \mgfe serve as the \dg criteria. For other elements\footnote{[Ba/Fe], [Cr/Fe], [Eu/Fe], [K/Fe], [La/Fe], [Na/Fe], [Nd/Fe], [Ni/Fe], [O/Fe], [Sc/Fe], [Si/Fe], [Ti/Fe], [Y/Fe], [Zn/Fe], [Zr/Fe]}, we see differences in the mean of the abundance distributions that marginally exceed the error on the mean. However, overall these are still extremely small, and all are below 0.05 dex. Note this difference is lower than the typical precision on these elements in individual stars.

\begin{figure*}[t!]
    \centering
    \includegraphics[width=0.95\textwidth]{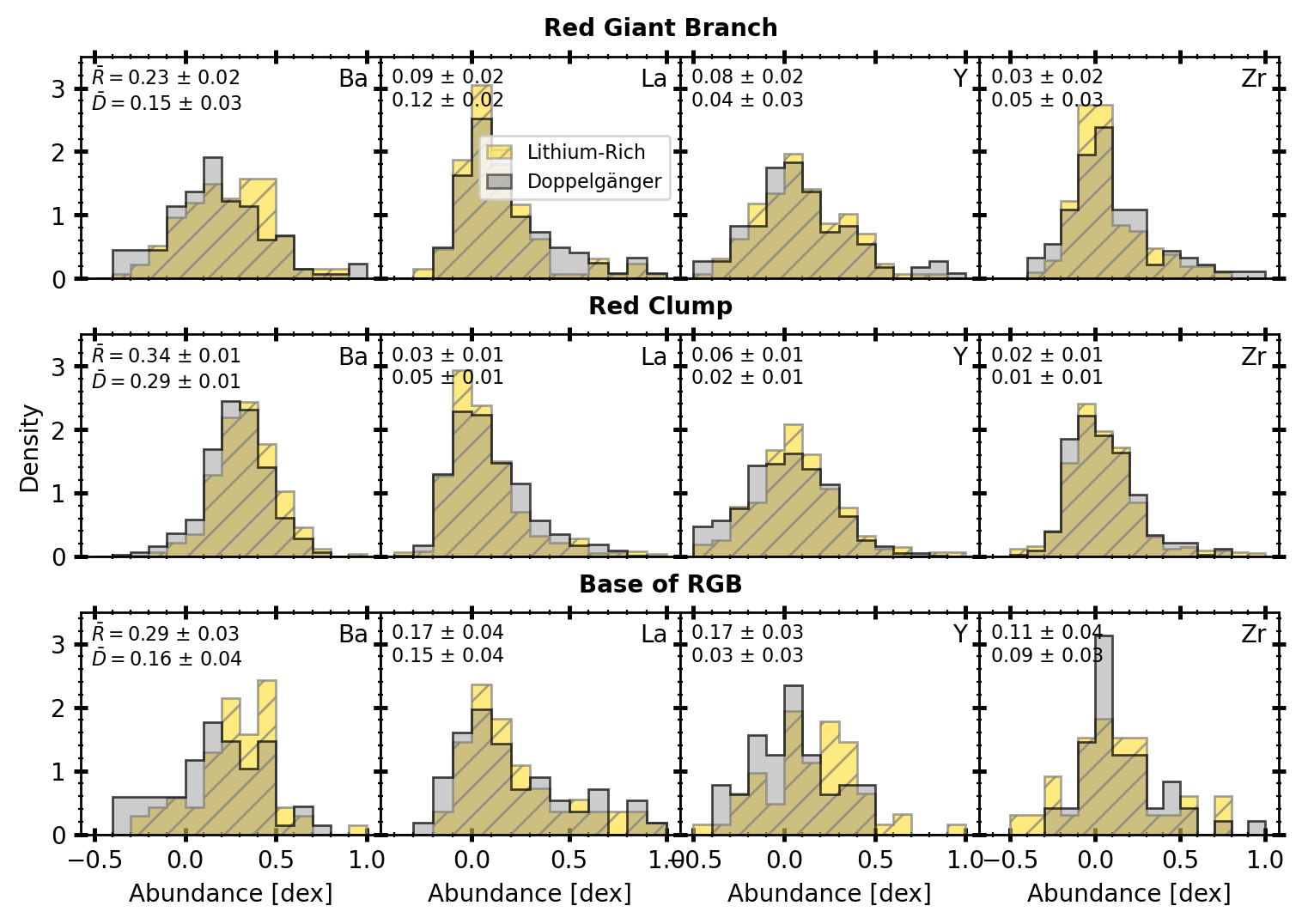}
    \caption{Distribution of \sprocess elements -- Barium (Ba), Lanthanum (La), Yttrium (Y), and Zirconium (Zr) -- for stars in three evolutionary states, namely red giant branch (top row), red clump (middle row), and base of the red giant branch (bottom row). The yellow and grey bars indicate \lirich and \dg stars, respectively, and the text indicates the median abundance for the \lirich (top) and \dg sample (bottom), and its associated error ($\sigma/\sqrt{N}$). There is no significant difference in all four elements for stars in the red giant branch and red clump phases; however, \lirich stars at the base of RGB have higher amounts of all four elements. This suggests that stars at the base of RGB are enriched in both lithium and \sprocess elements, likely by a companion intermediate-mass AGB star.}
    \label{fig:sprocess}
\end{figure*}
 
We reorganize the information in Figure \ref{fig:delta_abundance}, and show in Figure \ref{fig:diff_in_abundance} the average difference in each abundance, sorted in amplitude along the x-axis. While Figure \ref{fig:delta_abundance} includes all stars with reported abundances, where the histograms each show on the order of \app500 stars, Figure \ref{fig:diff_in_abundance}, which reports the mean difference in each abundance of the \lirich star and its \dg, includes \app15\% fewer stars since not all pairs have reported abundances for \textit{both} the \lirich star and its \dg for a given element. Therefore, the mean difference of the pairs may not be precisely the same as the difference on the means of the histograms in Figure \ref{fig:delta_abundance}, due to the samples being slightly different, and this now being a more robust test of differences in the populations. In Figure \ref{fig:diff_in_abundance}, we also exclude elements whose distributions were highly non-Gaussian as seen in Figure \ref{fig:delta_abundance}, and those where more than 90\% of the sample did not have good quality flags (e.g., Mo, Rb, Ru and Sr).

For Figure \ref{fig:diff_in_abundance}, we bootstrap the comparison of distributions of each element as follows: for each element abundance measurement and every \lirich star, we select a \dg at random from the 100 closest \dgs, and also select two additional \dgs. For every element, we find the mean difference between all \lirich and their first \dg, as well as the mean difference between two other \dgs of each \lirich star. We perform these steps for each element repeated 10 times, and use the average of 10 samples -- where each sample comprises a comparison of about \app500 stars for each element -- in Figure \ref{fig:diff_in_abundance}. For a given element, X, this can be formulated as follows,

\begin{equation}
   \Delta [\mathrm{X/Fe}] = \frac{1}{10} \sum_{j=1}^{10} \Bigg( \frac{1}{N}\sum_{n=1}^{N} {{X_{n}}} - {{X_{n'}}} \Bigg)
\end{equation}
where $j$ is the sample iteration (between 1-10), $N$ is the total number of stars used for the calculation for an element (varies for each element, since an abundance is not always available for both the Li-rich star and its \dg), $X_{n}$ and $X_{n'}$ are the abundances for the \lirich star and its first \dg respectively, or if analysing the reference set then two random \dgs for the Li-rich star. For each element, this gives us a bootstrapped comparison of the abundance distributions of Li-rich stars compared to 10 \dg distributions, and a reference sample of 10 \dg-\dg distributions.

The results are shown in Figure \ref{fig:diff_in_abundance}, where the yellow points show the mean difference between the \lirich and \lin pairs, and the grey points show the mean difference between pairs of \dgs of the \lirich stars. The error bars represent the 1$-\sigma$ dispersion of the 10 draws. The \dg-\dg difference measurement serves as a reference. The expectation is that these should show zero mean differences, as they are unbiased in being drawn from the same population in \teff, \logg, \feh, and \mgfe. Conversely, the mean differences in the \lirich and \lin pairs test if the bias in lithium between the pairs is associated with any difference in the individual elements.

We validate that \feh and \mgfe have a zero mean difference in the \lirich and \lin pairs, as expected for the two populations, since these parameters define the samples (e.g., see Figure \ref{fig:summary} for a summary). About \app60\% of the elements have differences in the \lirich-\lin pairs above 0.01 dex, but for \dgs-\dgs pairs, all elements show negligible differences, within the sampling uncertainties. The mean overall absolute difference of the \dg-\dg pairs is $\sim$0.0002 $\pm$ 0.0002 dex, but that of the \lirich-\dg pairs is $\sim$0.0200 $\pm$ 0.0002 dex, where the errors on these measurements is the inverse weighted variance. The reference test in Figure \ref{fig:diff_in_abundance} highlights the significant differences in abundances between \lirich and \lin pairs.

Despite having large samples of stars for each element, there are in some cases reasonably large 1$-\sigma$ standard deviation values of the 10 samples, for the comparisons between the \lirich-\dg and \dg-\dg distributions (e.g. V, Sm). This implies that there are abundance outliers driven by error on the measurements that bias these calculations in some cases, but which we mitigate the effect of with bootstrapping \citep[e.g., as seen in][]{Griffith_2022}.

\subsubsection{\sprocess Elements}

In our element abundance analysis, we see notable differences in the distribution of \sprocess elements. These represent an independent nucleosynthetic channel from our \dg criterion. Differences in the \sprocess elements may indicate a role of evolutionary state in the mechanism for Li-enrichment, or else transfer of \sprocess material from a binary companion. Stars with anomalously high \sprocess enhancements in particular, are proposed to be enhanced via mass-transfer from an AGB binary companion \citep[e.g.,][]{Cseh_2018, Norfolk_2019, Cseh_2022, denHartogh_2022, Escorza_2023}.

In Figure \ref{fig:sprocess}, we compare the \lirich and \lin distribution of four \sprocess elements for three evolutionary states. The RGB and RC phases are defined using the criteria in \cite{martell_2021}, and stars at the base of RGB have \logg $\geq 2.70$ dex (and less than \logg$=3.2$ dex given our sample construction, see Section \ref{sec:evol_state}). We compare the abundance distributions ([X/Fe]) of four elements, from left to right: Ba, La, Y, and Zr (indicated in the top right of each sub-panel). The text indicates the median and the error on the distribution ($\sigma\sqrt{N}$). Barium shows the largest difference in the median for all three evolutionary states, with $\Delta \sim 0.05-0.08$ dex higher in the \lirich sample. Li-rich stars at the base of RGB in particular (bottom row) have slightly higher abundances of \sprocess elements compared to their \dgs with mean differences exceeding the uncertainties, suggesting that \lirich stars at the base of the RGB are more likely to be \sprocess enhanced. 

Given the significant difference in Barium across all evolutionary states, we compared Barium between the \lirich and \dg sample. If we draw \dgs from a population with Li-measurements (and not upper limits), we see a secondary group of Barium-rich stars in the \dg sample that are otherwise missing from the \lirich sample. While the source of this result is inconclusive, we discuss this unusual finding in more detail in Appendix \ref{sec:app_dearth_barium}.

\subsection{Condensation Temperature Trends} \label{sec:cond_temp}

We also investigate the chemical abundances of our \lirich stars and their \dgs as a function of condensation temperature. Condensation temperature, \tc, is the temperature at which 50\% of an element condenses from gaseous to solid phase under protoplanetary conditions \citep{Lodders2003}. Analysis of refractory depletion and condensation temperature is widely used to probe dust and rock formation \citep[e.g.,][]{Venn1990, Savage1996, Heiter_2002, Maas_2005}. Given that rocky planets would form primarily out of high condensation temperature elements, the absence of these elements in the present-day stellar atmospheres has been suggested to arise from a history of planet formation \citep[e.g.,][]{Melendez2009, chambers2010}. Alternatively, engulfment of planetary material at late times in the formation process may give rise to enhancements in the abundances of high condensation temperature elements \citep[e.g.,][]{melendez2017, spinda2021}. Since this accretion of material would also enhance lithium abundance, we explore this explanation by searching our Li-rich stars for refractory material enhancements \citep[e.g.,][]{Ramirez2009, Gonzalez2010, Schuler2011, Ramirez2011, Melendez2012, Liu2014, Nissen2015, Schuler2015, Teske2016, Bedell2018, Liu2020, Nissen2020}.

\begin{figure*}[t!]
    \centering
    \includegraphics[width=1.\textwidth]{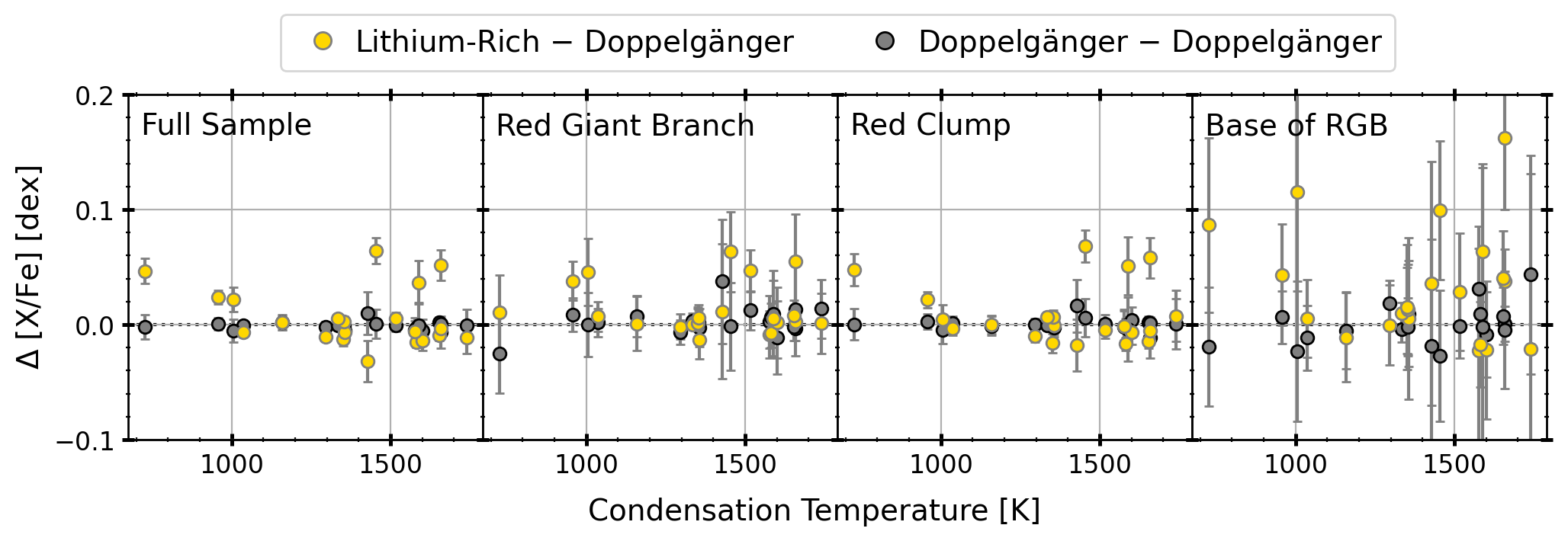}
    \caption{Similar to Figure \ref{fig:diff_in_abundance} organized by condensation temperature and separated into full sample, and three evolutionary states, indicated in the top left of each panel.}
    \label{fig:cond_temp_three_states}
\end{figure*}

To look for signatures of planets in our \lirich stars, we compare [X/Fe] of \lirich stars and their \dgs as a function of \cond for three evolutionary states, namely RGB, RC and base of RGB, as well as the full sample shown in Figure \ref{fig:cond_temp_three_states}. We limit the sample by only using stars with good quality flags for a given element. For all three stages, we found no significant difference in trend of [X/Fe] as a function of \cond, as can be seen in Figure \ref{fig:cond_temp_three_states} where the $\Delta$ [X/Fe] between the \lirich is similar to that of two random \dgs of the \lirich star. To ascertain that the trend is non-negligible for \lirich stars, we also analyzed the difference in abundance for two other \dgs for each \lirich star, but found no significant difference between the \lirich-\dg and \dg-\dg samples. 

A positive trend in [X/Fe] as a function of \cond would imply an excess of refractory elements in the \lirich star, while a negative slope would imply a dearth of refractory elements. Therefore, we compared the linear slope distribution for individual stars in the \lirich and \lin samples. We fit all elements with a condensation temperature above 1300 K on a [X/Fe]$-$\tc parameter space; this condition was chosen since \cite{Bedell2018} found a bias in the refractory abundance vs. \tc slopes if volatile elements (ie. \tc $< 1300$ K) were included. We performed bootstrapping to check if one element had a larger effect on the resulting trend. Removing [Na/Fe] and [Sc/Fe] had the most impact on the slope; however, the effect was insignificant and within $5-\sigma$ of the mean. We found a close overlap between the slope distributions of the \lirich and \lin samples. Our resulting distributions are similar to those found by \cite{Bedell2018} and \cite{Nibauer2021}, and we achieve a similar median to both. 

\section{Discussion}
\label{sec:discussion}
We have three primary results summarized below:
\begin{enumerate}[i)]
    \setlength\itemsep{0em}
    \setlength\parskip{0em}
    \item Section \ref{sec:dis_rc_vs_rgb}: \lirich stars are more likely to be red clump rather than  red giant branch stars. The increasing prevalence of red clump membership at higher Li-enrichment implies there is an evolutionary state dependence on lithium production, such as one associated with the He-flash. 
    \item Section \ref{sec:dis_halpha}: We detect differences in the \halpha in a subset of \lirich stars ($\sim$20\%), which is twice the incidence of the same signal in the \dgs. This could reflect potential differences in rotation, chromospheric activity, and/or binarity between \lirich stars and their \dgs. We also report a faster rotation for a subset of the \lirich\ population as measured by the broadening velocity parameter. This difference in the populations of \lirich and \dgs\ as seen in the spectra and broadening velocity indicates internal enrichment via binarity -- as a driver of the Cameron-Fowler mechanism -- is responsible for Li-enrichment, for a subset of the population. Whether the line profiles and broadening velocity are due to true differences in rotation, or if they are a direct binary detection, this profile links to the role of a companion for these stars.
    \item Sections \ref{sec:dis_cond_temp}: We see some population differences in the abundance distributions of the \lirich and \lin stars which change with evolutionary state, suggesting multiple mechanisms for Li-enrichment. In particular, \lirich stars show higher \sprocess abundances than their \dgs at the base of RGB, where internal mechanisms should not be responsible for Li-enrichment. This is a possible signature of mass-transfer from \textit{intermediate}$-$mass AGB companions, leading to both lithium and \sprocess enhancements. The lack of any condensation temperature trends and differences in refractory element abundances indicates no lines of evidence for the role of planetary ingestion.
\end{enumerate}

\subsection{Lithium in red clump and red giant branch stars}
\label{sec:dis_rc_vs_rgb}

We find a clear increasing probability of red clump membership as a function of Li-enrichment, in agreement with previous studies \citep[e.g.,][]{casey_2019, deepak_reddy_2019, deepak_2021_jul, deepak_2021_oct, martell_2021}. \cite{singh_2019} used \seismo to determine that most stars in their sample were from the He-core burning phase, similar to results found by \cite{Ming-hao_2021}. Similarly, \cite{zhou_2022} found that \app71\% of their \lirich sample belong to the red clump, and concluded that \lirich are rare in the red giant phase, while \cite{yan_2021} found a ratio of 75\% for RC to RGB stars (given \ali = 1.5 dex). 

Recent studies suggest a universal \li production mechanism occurs between the RGB and RC phases, potentially at the He-core flash. This has been proposed using evidence from observations \citep[e.g.,][]{kirby_2012, kirby_2016, kumar_2020, Mallick_2023}, and expectations from theoretical models \citep[e.g.,][]{Schwab_2020, Mori2021, Magrini2021}. However, \cite{chaname_2022} strongly argued there is no evidence for a lithium production event on the red clump when the dependency of lithium depletion on stellar mass in standard stellar models is accounted for, and that the results from \cite{kumar_2020} are biased due to sample selection. In fact, \cite{kirby_2012} attribute the brevity of this ubiquitous \li production event to the observed low number of \lirich giants. Our finding of the increasing prevalence of red clump membership for higher Li-enriched stars implies the He-flash induced lithium production is a plausible explanation for these observations. 

\subsection{Signatures of possible binarity}
\label{sec:dis_halpha}

In analyzing spectra of \lirich stars and their \dgs, stars selected to have the same stellar parameters (\teff, \logg, \feh, \mgfe) show similar, but non-identical spectra, where the most significant deviations are in the \halpha feature (see Figure \ref{fig:eg_spectrum}). We classified the differences at the H-$\alpha$ feature between \lirich\ and \dgs into $W$ and $M$ profiles, and found \app20\% of the \lirich\ population to have a $W-$designation compared to \app2\% of the \dgs. Section \ref{sec:flag32} shows that this profile is consistent with one of the stars in the pair being in a (spectroscopic) binary system. If all profiles directly tag (in effect, spectroscopic) binaries, this means the \lirich stars have tenfold the binary fraction of the reference population of \dgs for these architectures. 

We investigate the velocity broadening parameter (\vbr), as measured from the spectra. This confirms that the subset of \lirich stars with \halpha profiles that are indicative of faster rotation than their \dgs ($W-$designations) have preferentially higher broadening velocity than those with slower rotation than their \dgs ($M-$designations). This could be a true velocity broadening or simply due to the system being a binary and not single star. Nevertheless, this also showcases the differences in the \lirich and \dg population. 

As seen in Figure \ref{fig:vbroad_sarah}, there is a marked higher occurrence of stars with larger \vbr ($\gtrsim$ 20 \kms) for stars with \ali between $1.5-2.7$ dex, which is larger than the expected average \vbr (\app 10 \kms) for red giants \citep[e.g.,][]{Carney2008, Patton_2023}. Stars above this lithium threshold extend beyond 20 \kms, but a subset extend to \app50 \kms below it. Figure \ref{fig:vbroad_sarah} suggests there are likely multiple architectures leading to \lirich and super \lirich stars. The lesser \lirich stars appear to comprise the subset of the fastest rotators, while very few of the more enhanced stars are measured to be rotating faster than their \dg. This may be linked to the close binary fraction and detection limits of different types of spectroscopic binaries. 

The differences in the line profiles that we see in the H-$\alpha$ and Ca-triplet features could also be produced by differences in the magnetic activity levels on \lirich\ and \dgs\ stars.  \cite{Kowkabany_2022} found variable emission in the wings of the \halpha absorption line in the multi-epoch spectra of a recently discovered ultra \lirich metal-poor star. The authors associate this with a mass-loss event and possible outflows. An excess of magnetic activity is expected in both the presence of a binary companion \citep[e.g.,][]{montes_1996, sahai_2008} as well as planetary engulfment, since binary companions would lead to enhanced rotation in particular configurations due to tidal interactions which spin-up the primary \citep{casey_2019}. Therefore, it would be prudent to follow up the \galah \lirich targets similarly, with multi-epoch photometry and spectra, and/or radial velocity measurements from other surveys. This would test the role of binarity, rotation, and variability in the magnetic activity of the \lirich sample over time. 

In summary, the line profile and broadening velocity parameter results are clear lines of evidence for the role of binaries leading to \lirich giants. Binary systems are the main prediction for triggering the internal Cameron-Fowler mechanism of Li-enrichment. In this scenario, the primary in a binary system is tidally spun-up from the companion, which is proposed to induce this internal lithium production process \citep{casey_2019}. 

A particularly interesting and telling result that we uncover is the marginal differences in the mean abundances for some elements between \lirich stars and their \dgs. In particular, we see differences between the \sprocess abundances at the base of RGB, where the \lirich stars show higher \sprocess elements than their \dgs (see Figure \ref{fig:sprocess}). While we note that the differences are small, we see significant difference in [Ba/Fe] across all evolutionary states. Unlike along the RGB, any element abundance differences at base of RGB are unlikely to come from internal source \citep{gomez_2016a}, and therefore suggest an external mechanism of enrichment for these stars. These observations may be explained by intermediate-mass AGB companions \citep[$\gtrsim 4-8$ \solmass;][]{Uttenthaler_2007}.

Intermediate-mass AGB stars are theorized to make lithium in their envelopes via the process of `Hot Bottom Burning' \citep[HBB;][]{SackmannBoothroyd1992}. When the bottom of the convective layer reaches 40 MK, non-negligible nucleosynthesis can take place when the bottom of the convective layer merges with the outer hydrogen burning layers. HBB has been attributed to produce large amounts of \li found in the surfaces of evolved stars, especially those with a high \li abundance (\ali $\sim 4.5$ dex). If an AGB and RGB star are in a binary system and the AGB is Li-enhanced, then the AGB star could transfer some of its Li-enriched material to the red giant creating a super \lirich red giant. Since AGB stars are enriched in \sprocess elements, we would expect the red giant to also be similarly enriched \citep{SmithLambert1989}. 

Regardless, if the \lirich stars underwent a \plen event, accreted material from an AGB, or experienced rapid rotation caused by interaction with a binary or via tidal spin-up, we might expect to see an effect on chromospheric activity \citep[e.g.,][]{Metzger2012}. Therefore, we looked for emission in the UV using \galex data and emission in the IR using \wise, but found no differences between the \lirich and \lin populations in the two bands. Many studies have already utilized \galex to probe stellar activity \citep[e.g.,][]{Findeisen_2011, Shkolnik_2011, Stelzer_2016, dixon_2020}. However, we see no such UV excess for \lirich stars (see Figure \ref{fig:uv}). Finding no evidence of IR and UV emission could be due to observational bias, since only the brightest red giants would show high UV emission. In fact, \cite{Findeisen_2011} found little correlation between near-UV emission and activity for stars older than \app$0.5-1$ Gyr. Similarly, IR excess is mostly expected for young stellar objects.
\subsubsection{Binarity flags in \galah}
\label{sec:flag32}

\begin{figure}[t!]
    \centering
    \includegraphics[width=0.5\textwidth]{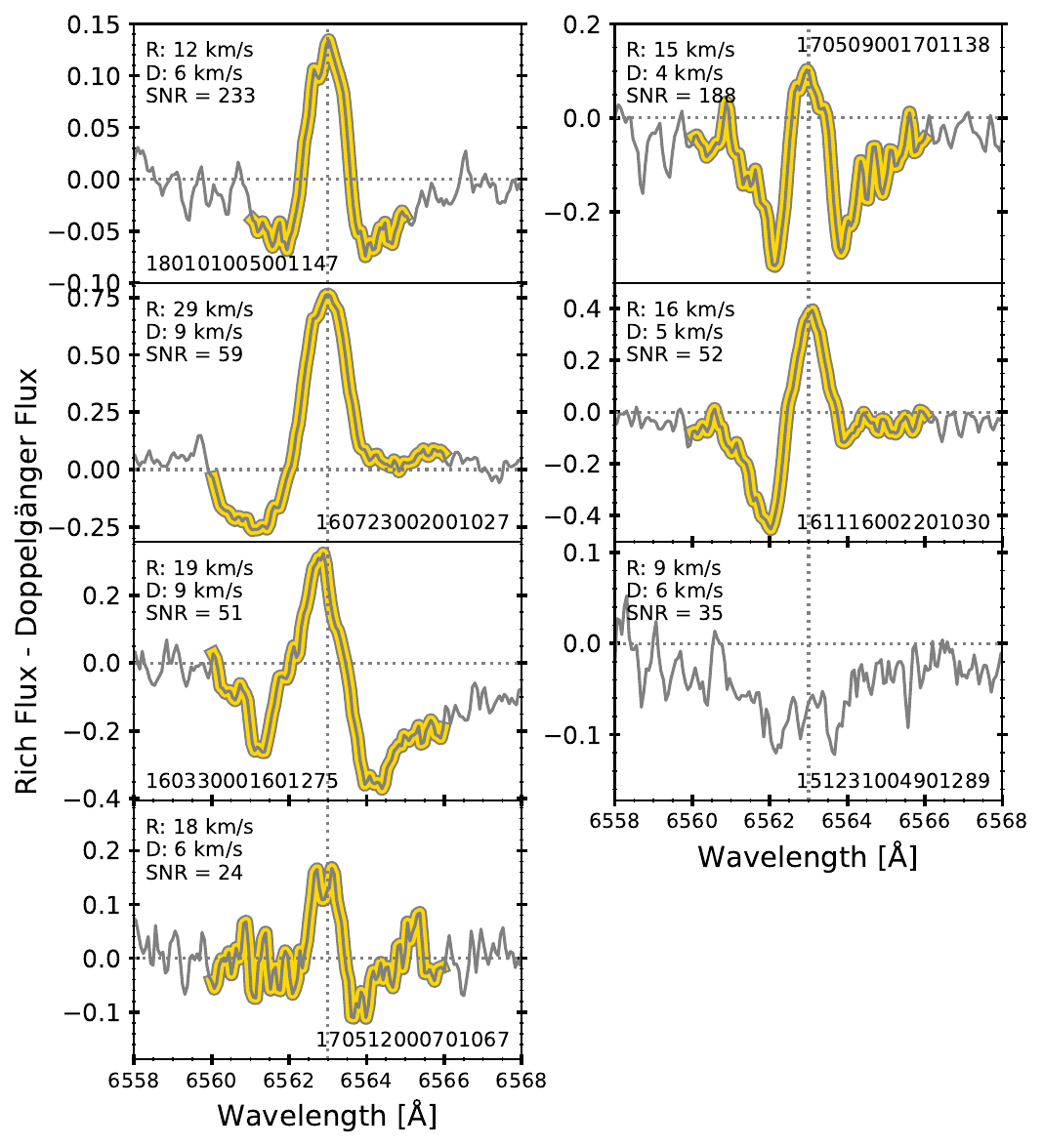}
    \caption{Difference in flux for 7 \lirich stars flagged as line-splitting binaries in \galah, centered on the \halpha line. The yellow curve highlights the $W-$profile, and the text includes the \vbr for both the \lirich star and its \dg, and the signal-to-noise of the \lirich spectrum in \galah.}
    \label{fig:flag32}
\end{figure}

During our sample construction, we require that all stars in \galah have \texttt{flag\_sp $\leq$ 1}, which \galah attributes to objects if no problems were identified in determination of their stellar parameters. However, \galah also flags line-splitting binaries, which we excluded in the original sample, when imposing the condition \texttt{flag\_sp $\leq$ 1}. Therefore to probe binarity, we now specifically analyzed these excluded, flagged stars. A \texttt{flag\_sp = 32} is given to a star by \galah if its spectrum looks similar to a line-splitting binary spectrum as found by a spectrum comparison algorithm (``tSNE"). Incorporating this condition results in 7 new \lirich stars, but no \lin stars. For the 7 new \lirich star, we find a \dg using the method described in Section \ref{sample_cons}, but without requiring that the difference in stellar parameters be less than the error on said parameter; this is justified since stellar parameters and abundances can be inaccurate for binaries. 

In Figure \ref{fig:flag32}, we plot the difference in flux for the 7 new \lirich stars and their \dgs, centered on the \halpha line, and include the \vbr for both stars. We notice that 6 of the 7 \lirich stars produce a $W-$profile with respect to their \dg, and even see a tentative $W-$profile in the seventh star, possibly less apparent due to the SNR of the \lirich spectrum, which is only 24. We also find that \vbr for the majority (5/7) of \lirich stars in Figure \ref{fig:flag32} is double the $v_\textrm{broad}$ of the \dgs.

The amplitude of these profile differences is extremely high in some cases, much higher than the \app 10\% level we detect in Figure \ref{fig:w_vs_m}. However, this clearly demonstrates that what we may be detecting with our $W$ and $M$ profiles is binary systems. In the case of the \lirich star being in the binary, the profile of the difference is the $W$ shape and when it is the \dg, it is the $M$. Subsequently, some of the enhanced rotation that is being measured and documented in Section \ref{sec:vbroad}, by the broadening velocity, may in fact be due to binarity being captured by this parameter. Therefore, we may have created, via the \dg reference comparison, a highly sensitive method to detect spectroscopic binaries that are otherwise evading detection. These stars will not be flagged binaries with the RUWE parameter, which is instead preferentially sensitive to wide separation binaries (discussed in detail in Section \ref{sec:paired}). Indeed, the 6 of the 7 stars shown in Figure \ref{fig:flag32} all have RUWE $<$ 1.2. However, these stars would presumably show radial velocity variations over time detectable in multi-epoch survey data. We intend to follow up the stars that we have flagged as either $W$ and $M$ with complementary survey data to verify what fraction of these are indeed in binary systems.

\subsubsection{Expected parameter space of potential binary systems}

\begin{figure}[t!]
    \centering
    \includegraphics[width=0.5\textwidth]{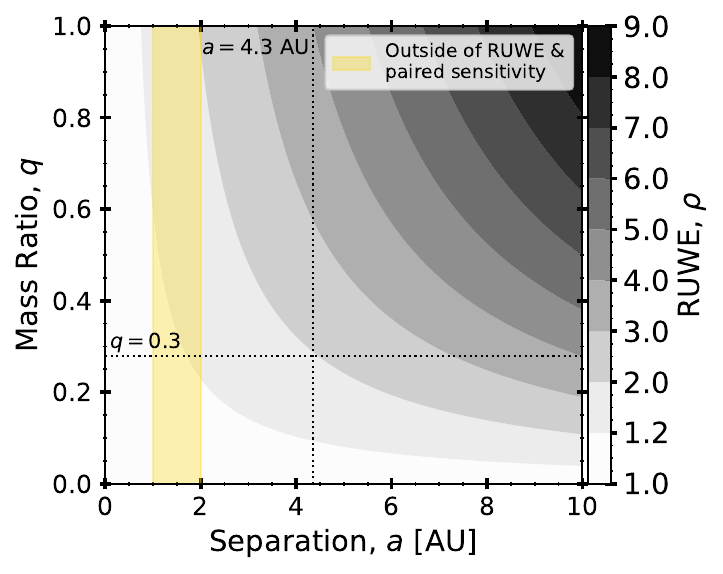}
    \caption{Expected RUWE values for a binary system given a range of orbital separations and mass ratios. Given the observed RUWE values, we can eliminate regions of this parameter space where RUWE $\gtrsim 4$. The dashed lines are shown as reference for relevant boundaries of the mass ratio and separation.}
    \label{fig:ruwe_plot}
\end{figure}
Although RUWE values derived in \gaia can be used to identify binary architectures, we can also predict the separations and mass ratios that would likely produce the observed RUWE values in our sample. An unresolved binary would cause a single-source astrometric model to perform poorly, resulting in a high \chisq; this is equivalent to RUWE or $\rho$, the re-normalised unit weight error. We derive $\rho$ for stars with orbital separations between $0-10$ AU, and mass ratios between 0 to 1. We briefly outline the steps below, and direct the reader to \cite{Belokurov_2020} for a more detail description of the methods. We calculate $\rho$ using the following,

\begin{eqnarray}
\delta \theta \; &=& \; \sqrt{<\delta \theta^2_i>} \; \approx \;\sigma_{\textrm{AL}}(G) \sqrt{\rho^2-1} \\
\frac{\delta a}{\textrm{AU}} \; &=&  \;\frac{\delta \theta}{\textrm{mas}} \frac{D}{\textrm{kpc}}\\
\delta a \; &\propto& \;\frac{a|q-l|}{(q+1)(l+1)}
\label{ruwe_eq}
\end{eqnarray}

given a mass ratio $q=m_2/m_1$, luminosity ratio $l=l_2/l_1$, angular perturbation $\delta \theta$, wobble $\delta a$, and distance to source $D$. We assume $l=0$ since we cannot see the secondary companion for our stars, and $D=2.2$ kpc which is  the mean distance for our sample. Given a $G$ magnitude, we use Figure 9 in \cite{Lindegren_2018} to approximate $\sigma_{\textrm{AL}}$, the per-scan along-scan centroiding error. The mean $G$ magnitude of our stars is 12.2 which results in $\sigma_{\textrm{AL}}=0.25$. 

In Figure \ref{fig:ruwe_plot}, we show the predicted RUWE for a range of orbital separations and mass ratios. As expected, the largest RUWE is seen at the largest mass ratio, since the higher the mass of the faint object, the more the wobble of the bright object ($l_1$, $m_1$). Similarly, RUWE is higher for large separations since the wobble produced by the star on the sky is larger. For the range of RUWE values in our sample, which is $0.6-4$, the parameter space that covers these RUWE values are, for the majority, stars at close separations, of  $a\sim0-6$ AU, for a range of mass ratios between companions. We exclude the presence of any binaries along the darkest coloured contours, i.e. at separations $\gtrsim$ 7 AU at mass fractions $\gtrsim$ 0.5. If the \lirich stars are in binaries, the RUWE values demonstrate that these would be close enough for past mass-transfer between the primary and an AGB companion, where a mass-transfer event could enhance the primary in lithium and other elements (ie. \sprocess elements). 

\begin{figure}[t!]
    \centering
    \includegraphics[width=0.5\textwidth]{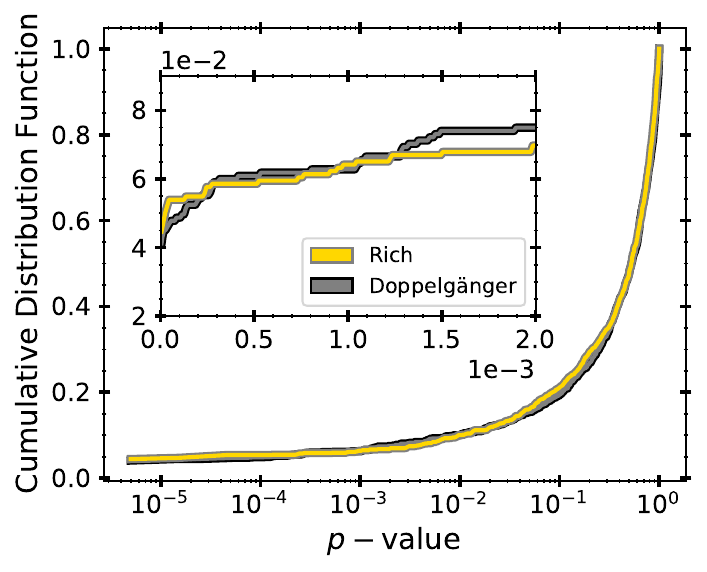}
    \caption{Cumulative distribution function (CDF) of $p-$value of \lirich and \lin stars in \paired catalog where both the \lirich star and its \dg have a $p-$value available. The value of CDF at $p-$value $= 0.001$ corresponds to stars flagged as binaries. The CDF of the two samples is similar which suggests that either the \lirich in binaries have larger separations than the method is sensitive to detect ($>$ a few AU), or another mechanism causes Li-enrichment for these stars. The inset plot shows the zoomed-in distribution around $p-$value $= 0.001$.}
    \label{fig:cdf}
\end{figure}

\begin{deluxetable*}{c||cc|cc}
\tablecaption{Analysis with \paired for \lirich stars and their \dgs. Stars with a RUWE $\geq$ 1.2 are predicted to be contaminated by a close companion, while stars with $p-$value $\leq$ 0.001 are flagged as potential binaries in \paired.\label{tb:paired}}
\tablehead{\colhead{} & \multicolumn{2}{c}{Li-rich}  & \multicolumn{2}{c}{Doppelg\"anger} } 
\startdata
Total & 1099 & $-$ &1099 & $-$\\
In \paired & 1075/1099 & (98 $\pm$ 0.4\%) &1067/1099 & (97 $\pm$ 0.5\%) \\ 
RUWE $< 1.2$ & 1011/1075 & (94 $\pm$ 0.4\%) &  981/1067 & (92 $\pm$ 0.8\%)\\
RUWE $\geq 1.2$ & 64/1075 & (6 $\pm$ 0.7\%) &  86/1067 & (8 $\pm$ 0.8\%)\\
$p-$val $\leq 0.001$ & 69/1075 & (6 $\pm$ 0.7\%) & 67/1067  & (6 $\pm$ 0.7\%) \\
RUWE $< 1.2$ \& $p-$val $\leq 0.001$ & 60/1011 & (6 $\pm$ 0.7\%) & 61/981  & (6 $\pm$ 0.8\%)\\
RUWE $\geq 1.2$ \& $p-$val $\leq 0.001$ & 9/64 & (14 $\pm$ 4\%) & 6/86 & (7 $\pm$ 3\%) \\
\enddata
\end{deluxetable*}

\subsubsection{Evidence of binarity with \paired}
\label{sec:paired}

To search for more direct evidence of stellar multiplicity, we used \paired, a statistical framework that uses \gaia radial velocity error measurements to search for binarity \citep{chance_2022}, which typically detects binaries at separations of up to a few AU and mass ratios above 0.1. Stellar multiplicity can be detected in the \gaia radial velocity measurements as the presence of excess noise, compared to stars of similar apparent magnitude and colour. With proper calibration of the expected radial velocity jitter for similar sources, one can estimate a probability that a source is a non-single star based on its reported radial velocity error in \gaia, and the number of radial velocity transits. 

We analyzed the \paired probability for our sample in combination with RUWE which is sensitive to close binaries, where stars with RUWE $\geq 1.2$ are likely to be binaries \citep[e.g.,][]{berger_2020}. For both the \lirich and \lin samples, we compare the number of binaries based on the RUWE and \paired criteria, where a $p-$value $\leq 0.001$ is flagged as a potential binary in \paired. Table \ref{tb:paired} shows the results of this comparison where the errors on the percentages represent the uncertainty in a Bernoulli trial. 

Of the \lirich stars that have a RUWE $\geq$ 1.2, only 14\% (9/64) are classified as binaries based on RUWE and \paired. In contrast, of the \dgs with a RUWE $\geq$ 1.2, 7\% (6/86) are possible binaries. These low fractions are in agreement with Figure \ref{fig:ruwe_plot} that demonstrates that RUWE $\geq$ 1.2 arises from systems with $a \geq 2$ AU, which paired is not sensitive to. For stars with RUWE $< 1.2$, which corresponds to the majority of our sample, 6\% (60/1011) of the \lirich and 6\% (61/981) of the \dg sample are binary detections in \paired.

We suspect that $W-$designated \lirich stars and the $M-$designated \dg stars are in binary systems, that have relatively close separations ($\lesssim 2$ AU, see Figure \ref{fig:ruwe_plot}) and are up to near-equal mass (from the line difference profiles, see Figure \ref{fig:w_vs_m}). Therefore, we compared \paired classification for the \lirich stars with $W$ and $M$ designations. Of the $M-$designated \lirich stars, none are classified as binaries in \paired, compared to 11\% of the $W-$designated \lirich stars. Therefore, $W-$stars are substantially more likely to be binaries according to \paired, compared to $M-$stars (where the \dg would be the potential binary). This provides clear evidence to associate the $W$ and $M$ profiles with binary architectures. 

We also examined the \paired binary fraction as a function of evolutionary state for the \lirich and \lin samples. Of the 126 \lirich RGB stars, 15\% are classified as binaries in \paired compared to 5\% of the \lirich RC stars. Similarly, of the 139 \lin RGB stars, 14\% are classified as binaries, compared to 4\% of the \lin RC stars. Given the similar fractions, the two populations are the same in terms of binary architectures detected via RUWE and \paired.

In Figure \ref{fig:cdf}, we show the cumulative distribution function of $p-$values for \lirich stars and their \dgs. The distributions are marginally different at the threshold value of $p-$value $\leq 0.001$: the \lirich stars reach 6\% compared to 7\% for the \dgs. This implies that the selection function of \galah renders the recovery rate of binaries very small. 

In summary, the RUWE analysis rules out systems with mass ratios above 0.5 at separations of 7 AU, or mass ratios of \app 1 at separations of 5 AU, while \paired is sensitive to separations $< 1$ AU. We can see from Figure \ref{fig:ruwe_plot} that any binary systems with RUWE $< 1.2$ will be restricted to systems with separations of $\lesssim 2$ AU for equal mass binaries, and $\lesssim 3$ AU for binary mass ratios of \app 0.4. The \paired analysis shows the $W$ and $M$ designations in the spectra comparisons are associated with binarity, and that there is a higher overall binary fraction for the \lirich population compared to the \lin. However, the binary detection rate is extremely low overall.

\subsection{Sub-stellar object engulfment as a mode of enrichment on the RGB}

\label{sec:dis_cond_temp}
Accretion of a planet or brown dwarf is expected to raise abundances of surface elements, including \li \citep[e.g.,][]{Koch2011, Koch2012}. \cite{SiessLivio1999a, SiessLivio1999b} were the first to explore the \plen scenario as a possible mechanism for Li-enhancement, but many studies have suggested \plen as a possibility for Li-enhancement \citep[e.g.,][]{adamow_2012, gomez_2016a, gomez_2016b}. However, while some place an upper limit on the mass of the engulfed object that could produce an observable signature in the star \citep[e.g., 15 M$_\text{J}$ in][]{gomez_2016a}, others suggest no connection between lithium and the presence of planets \citep[e.g.,][]{Baumann2010}.

Indeed if \plen were the mechanism for a significant fraction of \lirich RGB stars, then we might see an excess in refractory elements for these targets, as well as trends in condensation temperature that differ between \lirich and \lin stars. Motivated by this idea, we derived the [X/Fe]$-$\tc slope for all stars in our \lirich and \dg samples, but found no difference in measurements between the two samples, nor an excess in refractory elements across evolutionary states; we therefore find no evidence for planetary engulfment in our study. Despite our null results, we do not believe \plen should be excluded as a possible mechanism for Li-enrichment for two reasons. 

Firstly, previous studies that compared \cond and chemical abundances in planet hosts are inconsistent in their findings, and there is a lack of consensus on whether \cond is an appropriate indicator for the presence of planet formation itself. For instance, \cite{Ramirez2011} and \cite{TucciMaia2014} studied the chemical composition as a function of \cond in 16 Cygni where 16 Cyg B is a planet host. They find no trend between abundance differences and \cond, nor any correlation between \cond and the planet host star. This is in agreement with other similar studies \citep[e.g.,][]{Liu2014, Teske2016, Liu2020, deepak_2020, Nissen2015}. On the contrary, some studies find promising results in this parameter space. For instance, \cite{Melendez2009} found that the Sun is more depleted in refractory elements as compared to solar-twins, with others producing similar results \citep[e.g.,][]{Ramirez2009, Gonzalez2010}. Interestingly, some studies even found contradicting results in their own sample \citep[][]{Schuler2011, Schuler2015}. 

Secondly, we must also consider timescales involved with \plen, such as lithium survival and planet inspiral timescales. Red giants are expected to remain \lirich for \app2 Myr \citep{casey_2019}, and even significant signatures from \plen would have a maximum Li-7 survival time of $0.5-0.9$ Gyr \citep{melinda_2021}. However, \cite{Behmard_2022} reported that \plen signatures are short lived, \app 90 Myr for \app 1 \solmass star, and that these signatures would no longer be observable \app 2 Gyr after engulfment. Considering that our \lirich stars range from $0.4 - 11.2$ Gyr old, where RGB stars are $\sim 1-11$ Gyr old, it may be difficult to observe the evidence of any engulfment in our sample, since engulfment signatures are difficult to see for older stars \citep[][]{Behmard_2022, Behmard_2022_twins}.

Assuming \plen were the sole mechanism for Li-enrichment, we can calculate the approximate expected number of \lirich RGB stars. From analysis done by \cite{melinda_2021}, signatures of engulfment are diluted at later stages of post main-sequence evolution, with abundance measurements falling below 1.5 dex. Therefore, to observe \li enrichment signature with statistical significance, the host star should have a mass of $1.4-1.6$ \solmass, with the strongest signature occurring for stars with mass above 1.4 \solmass. Of the 449,553 RGB stars in \galah, 21,113 have a stellar mass between $1.4-1.6$ \solmass; therefore, we would expect \app211 \lirich RGB stars, assuming the engulfed planet is a hot Jupiter and \app1\% of giants are expected to host these planets. In principle, all \nrgb RGB stars in our \lirich sample could be explained by planetary or brown dwarf engulfment, but realistically we expect only a fraction of \lirich RGB stars to have undergone \plen \citep[e.g.,][]{casey_2019, melinda_2021}.

\begin{deluxetable}{l|cc}
\tabletypesize{\scriptsize} 
\tablecaption{Quantitative summary of main results.\label{tb:results_summary}}
\tablehead{\colhead{Measurement} & \colhead{Li-rich} & \colhead{Doppelg\"anger}} 
\startdata
RUWE $\geq 1.2$ ($\gtrsim$ 2 AU separation binaries) & 6\% & 8\% \\
$p-$val $\leq 0.001$ ($\lesssim$ 1 AU separation binaries) & 6\% & 6\% \\
Spectroscopic binary flag set in GALAH & 7 stars & 0 stars \\
$W-$designation in \halpha & 20\% & 2\% \\
\vbr $\geq 20$ \kms & 2.6\% & 0.2\% \\
Renamed sources between \gaia DR2 \& DR3 & 0.4\% & 1.0\% \\\hline
\multicolumn{3}{c}{Results for Li-rich sample only}\\\hline\hline
\multicolumn{3}{c}{{Red clump fraction (\lirich:background})}\\
\multicolumn{1}{c|}{\ali $=1.5-1.8$ dex} & \multicolumn{2}{c}{1.4} \\
\multicolumn{1}{c|}{\ali $=2.8-4.2$ dex} & \multicolumn{2}{c}{2.2} \\
\multicolumn{3}{c}{{Mean elemental abundances relative to \dgs}} \rule{0pt}{3ex}\\
\multicolumn{1}{c|}{$\Delta\mathrm{(Ba)}$} & \multicolumn{2}{c}{0.06 $\pm$ 0.01}  \\
\multicolumn{1}{c|}{$\Delta\mathrm{(Y)}$}& \multicolumn{2}{c}{0.05 $\pm$ 0.01}  \\
\multicolumn{1}{c|}{$\Delta\mathrm{(O)}$} & \multicolumn{2}{c}{0.04 $\pm$ 0.01}  \\
\multicolumn{1}{c|}{$\Delta\mathrm{(Zn)}$} & \multicolumn{2}{c}{0.05 $\pm$ 0.01}  \\
\multicolumn{1}{c|}{$\Delta\mathrm{(Na)}$} & \multicolumn{2}{c}{0.02 $\pm$ 0.01}  \\
\multicolumn{3}{c}{{Median elemental abundances at the base of RGB}} \rule{0pt}{3ex}\\
\multicolumn{1}{c|}{Ba} & \multicolumn{2}{c}{0.29 $\pm$ 0.03} \\
\multicolumn{1}{c|}{La} & \multicolumn{2}{c}{0.17 $\pm$ 0.04} \\
\multicolumn{1}{c|}{Y} & \multicolumn{2}{c}{0.17 $\pm$ 0.03} \\
\multicolumn{1}{c|}{Zr} & \multicolumn{2}{c}{0.11 $\pm$ 0.04} \\
\enddata
\end{deluxetable}

\section{Conclusion}
\label{sec:conclusion}
We have undertaken a thorough analysis, with the aim to determine the formation mechanisms of \lirich giants, using new available data from \galah, and complementary data from \gaia DR3 and \galex. We assembled a sample of \total \lirich stars in \galah, and compared it to a sample of otherwise identical \lin star. Our main conclusions are as follows, quantitatively summarized in Table \ref{tb:results_summary}: 

\begin{enumerate}[a)]
    \setlength\itemsep{0em}
    \setlength\parskip{0em}
    \item Figures \ref{fig:delta_abundance}, \ref{fig:diff_in_abundance}, \ref{fig:sprocess}: inspection of \sprocess elements at varying evolutionary states suggests mass-transfer from either low-mass or intermediate-mass AGB companions. This findings suggests that for stars at the base of RGB, mass-transfer from an \textit{intermediate}$-$mass AGB leads to enrichment in both lithium and \sprocess elements.
    \item Figures \ref{fig:flag32}, \ref{fig:ruwe_plot}, \ref{fig:cdf}: examination of radial velocity error in \gaia in combination with simulated RUWE values suggests that binarity could be responsible for a subset of \lirich stars. This mechanism is likely initiated over a limited range of orbital parameters. The RUWE measurement excludes systems with mass ratios $\gtrsim 0.5$ at separations $\gtrsim 7$ AU. Conversely, \paired is only sensitive to binary separations of $< 1$ AU, and confirms that $W-$designated \lirich stars are preferentially binaries. The corresponding RUWE for these stars in combination with the \paired detection restricts their parameter space to separations of $\sim 1-2$ AU for equal mass binaries, and up to 1 AU for mass ratios of 0.5. These would allude detection as binaries in both \paired and RUWE, but may be showing up in the spectra in the two-fold incidence of $W-$profiles compared to $M-$profiles. Furthermore, stars flagged as line-splitting binaries in \galah are exclusively \lirich, and not \lin. These stars are 6/7 times classified as $W-$profiles, and have higher \vbr compared to their \dgs. This strongly suggests that $W$ and $M$ profiles are spectroscopic binaries at restricted separations.
    \item Figures \ref{fig:eg_spectrum}, \ref{fig:w_vs_m}, \ref{fig:vbroad_sarah}, \ref{fig:vbroad_dg}, \ref{fig:vbroad_diff}: differential analysis of \galah spectra show that a subset of \lirich stars are rotating faster than their \lin counterparts. This may be intrinsic rotation or the impact of binary systems on this parameter. We find twice as many anomalously high rotators (\vbr $\gtrsim 20$ \kms) in the \lirich sample compared to the \dg population. Higher broadening velocities are seen for stars below \ali $= 2.7$ dex.  The changing distribution of broadening velocities and $W-$designations for the \lirich population as a function of lithium implies multiple mechanisms of enrichment. 
    \item Figures \ref{fig:teff_vs_logg} \& \ref{fig:random_logg}: the increasing prevalence of red clump stars at higher lithium enhancement points to an event between the red giant branch and red clump phases of stellar evolution, or at the He-flash itself, that causes Li-enrichment.
    \item Figures \ref{fig:diff_in_abundance}, \ref{fig:cond_temp_three_states}: we find no evidence for the role of planetary engulfment from condensation temperature or refractory element trends. However, this finding does not rule out planetary engulfment as a possible scenario for Li-enrichment, since planetary engulfment signatures are short lived, and dependent on stellar age.
    \item Figure \ref{fig:uv}: we find no difference in UV and IR emission using \galex and \wise data, respectively, between populations of \lirich and \lin stars.
\end{enumerate}

Our analysis provides evidence of multiple mechanisms of Li-enrichment of red giants, including both internal and external modes. Our work provides direct evidence that a subset of \lirich stars are preferentially in binary systems, and a subset have likely undergone mass transfer from an intermediate-mass AGB companion; we report the first evidence for systematic differences in element abundances for the \lirich population compared to the field. We have also converged on a restricted parameter space for binarity for the majority of the binary architectures, using the complementary \gaia measurements for the \galah stars.

Lastly, we find increasing red clump membership for higher Li-enriched stars, which suggests a He-flash induced lithium production; however, the data do not differentiate between production \textit{by} the He-flash, or from tidal-locking triggered CF production from a companion on the red clump \citep[e.g,][]{casey_2019}. One outstanding issue with the former, as noted by \cite{casey_2019}, is the origin of the required He-3 reservoir and transportation of beryllium at temperatures where lithium could be created and persist post He-flash. Conversely, an issue with the latter is that we do not detect a substantial number of binary companions for the \lirich stars in excess of the reference \lin population. However, companions could be evading detection at separations just outside of and interior to the sensitivity of radial velocity and astronomy detection limits with \paired and RUWE, respectively. 

Further investigation of our sample, such as by measuring rotation periods from time-series observations, differentiating between intrinsically higher rotation and binarity, and simulating binary architectures, would enable us to model and differentiate between the roles of these different mechanisms that we are tapping into in more detail. 

\section{Acknowledgments}
We are thankful to our anonymous referee for helpful comments which improved this manuscript. MS would like to acknowledge the support of the Natural Sciences and Engineering Research Council of Canada (NSERC). Nous remercions le Conseil de recherches en sciences naturelles et en génie du Canada (CRSNG) de son soutien. MS thanks the LSSTC Data Science Fellowship Program, which is funded by LSSTC, NSF Cybertraining Grant \#1829740, the Brinson Foundation, and the Moore Foundation; her participation in the program has benefited this work. SLM acknowledges the support of the Australian Research Council through Discovery Project grant DP180101791 and the support of the UNSW Scientia Fellowship Program. This work was supported by the Australian Research Council Centre of Excellence for All Sky Astrophysics in 3 Dimensions (ASTRO 3D), through project number CE170100013. B.D.M. is supported in part by the National Science Foundation through the NSF-BSF program (grant number AST-2009255).  The Flatiron Institute is supported by the Simons Foundation.

This work made use of the Third Data Release of the \galah Survey \citep{galah_sven_2021}. The \galah Survey is based on data acquired through the Australian Astronomical Observatory, under programs: A/2013B/13 (The \galah pilot survey); A/2014A/25, A/2015A/19, A2017A/18 (The \galah survey phase 1); A2018A/18 (Open clusters with HERMES); A2019A/1 (Hierarchical star formation in Ori OB1); A2019A/15 (The \galah survey phase 2); A/2015B/19, A/2016A/22, A/2016B/10, A/2017B/16, A/2018B/15 (The HERMES-TESS program); and A/2015A/3, A/2015B/1, A/2015B/19, A/2016A/22, A/2016B/12, A/2017A/14 (The HERMES K2-follow-up program). We acknowledge the traditional owners of the land on which the AAT stands, the Gamilaraay people, and pay our respects to elders past and present. This paper includes data that has been provided by AAO Data Central (\url{datacentral.org.au}).

This work has made use of data from the European Space Agency (ESA) mission \gaia (\url{https://www.cosmos.esa.int/gaia}), processed by the \gaia Data Processing and Analysis Consortium (DPAC, \url{https://www.cosmos.esa.int/web/gaia/dpac/consortium}). Funding for the DPAC has been provided by national institutions, in particular the institutions participating in the \gaia Multilateral Agreement. This paper makes use of data products from the Wide-field Infrared Survey Explorer, which is a joint project of the University of California, Los Angeles, and the Jet Propulsion Laboratory/California Institute of Technology, funded by the National Aeronautics and Space Administration.

\facilities{
\gaia \citep[][]{gaia_collab, gaia_2018, gaiadr3},
\galah \citep[][]{galah_survey, galah_sven_2021},
\galex \citep[][]{galex_martin, galex_ais}
}

\software{astropy \citep{2013A&A...558A..33A,2018AJ....156..123A}, Matplotlib \citep{matplotlib}, NumPy \citep{harris2020array}, Pandas \citep{mckinney-proc-scipy-2010}, SciPy \citep{2020SciPy-NMeth}}

\appendix

\begin{figure}[t!]
    \centering
    \includegraphics[width=0.5\textwidth]{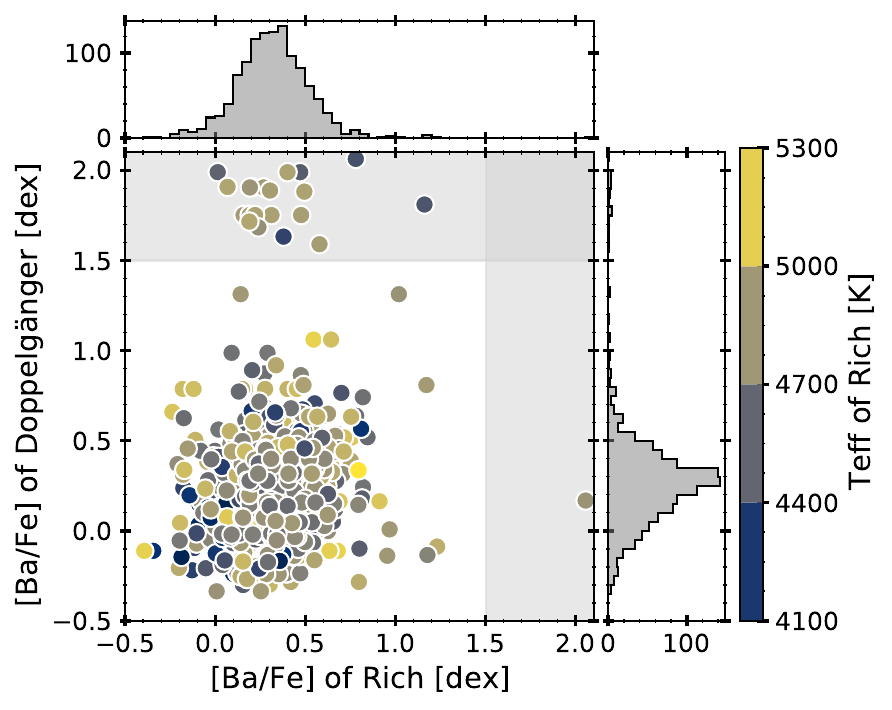}
    \caption{Distribution of [Ba/Fe] for \lirich and \lin samples coloured by \teff of \lirich stars. The highly-enriched barium stars that are seen in the \dgs are not present in the \lirich sample. This is evidence that the mechanism by which stars become Ba-rich, likely the mass-transfer of a close low-mass AGB companion, is incompatible with the mechanism for Li-enrichment. Note that two stars fall outside of the axis limits.}
    \label{fig:barium}
\end{figure}

\section{Reference sample comparison}\label{sec:app_upper_limits}

For this work, the \dgs were selected from a population with upper limits for [Li/Fe] (\texttt{flag\_Li\_fe==1}), and \ali $<$ 1.0 dex. However, we also undertook an analysis using an alternate reference set, by selecting \dgs with measured values for [Li/Fe] (\texttt{flag\_Li\_fe==0}), which resulted in 7543 stars from which to select \dgs. The majority of our results stayed the same except in two aspects.

First, using the \dgs with a Li-measurement, we find three times as many $M-$stars, but roughly the same number of $W-$stars, compared to the \dgs with upper limits for lithium. We expect that this lower fraction of $M-$stars using the upper-limit doppelganger set is a consequence of their lower SNR; the mean SNR for the upper limit sample of \dgs is $\sim$51, but the mean SNR of the \dg sample with Lithium measurements is $\sim$72. Our ocular classification of $M-$ and $W-$stars is SNR dependent -- a higher noise would obfuscate this.

Second, we find that the \dgs with a measurement of Li have an anomalously high fraction of Ba-rich stars, with [Ba/Fe] $>$ 1.5. This suggests that there is some correlation between the Lithium content in a star and mechanism for Ba-enrichment. This is discussed in Appendix B.


\section{Dearth of Barium-rich stars that are Lithium-rich} \label{sec:app_dearth_barium}

When we selected \dgs with a Lithium measurement as described in Appendix A and compared the distribution of \sprocess elements between the \lirich and \dg samples, we found a surprising result where a subset of Ba-rich stars that are seen in the \dg population are not present in the \lirich sample. This is shown in Figure \ref{fig:barium} where we see a small population of stars in the \dg sample with high Barium. We found a similar result for the three other \sprocess elements (La, Y, Zr) as we do using Barium. Further analysis revealed that this is a peculiarity of the \dg sample, and not the Li-rich stars. We found that for both the Li-rich stars and their \dgs (and their parent sample) with Lithium upper-limits, 0.2\% of stars have [Ba/Fe] $>$ 1.5 dex. However, for the \dgs set with Lithium measurements, we find that 1.4\% (including of the parent sample that the \dgs are drawn from) have [Ba/Fe] $>$ 1.5. We note in Appendix A that the two \dg sets (upper limit and measurement) have different SNR distributions. We therefore implemented an additional criterion for both \dgs analyses, whereby each reference object was required to have an SNR difference of $\leq$ 25 compared to each Li-rich star. Despite this additional criterion, the result whereby a substantially larger fraction of Ba-rich stars are present in the \dgs measurement sample still persisted. This implies that there is some relationship between the Li-abundance of a star and the ability of a star to become Ba-enriched.

Highly Ba-rich stars are believed to have been recipients of \sprocess material from a low-mass AGB companion via mass-transfer as the secondary overflows its Roche lobe \citep[e.g.,][]{Stancliffe_2021, Cseh_2022, Norfolk_2019}. Statistically, studies have shown that barium stars are found in binary systems \citep[e.g.,][]{mcclure1980, mcclure1983, jorissen1988, mcclure1990, Jorissen1998, Hansen_2016}, where the secondary would now be a white-dwarf in observed systems.

Nominally, the lower mass range of the AGB ($\leq$ 4 \solmass) does not produce lithium in HBB \citep[][]{Ventura_2020}. However, although low-mass AGB stars are the likely companion culprits of \sprocess enriched stars, this result demonstrates that there is some preferential window of measured Li-abundance (\ali $<$ 1 dex) where the Ba-rich stars are most prevalent compared to stars highly enriched in \li (\ali $>$ 1.5 dex), or the stars with upper limits of lithium only. Thermohaline mixing has been proposed as a possible alternative route via which low-mass AGB stars can produce lithium \citep{Cantiello_Langer_2010}. This result indicates that this mechanism may be present but is inefficient, or short lived, as these highly enriched Barium stars are not particularly \lirich, but may boost the Li-abundance above the mean of the population.

\begin{deluxetable*}{cccc}
\tablecaption{Survey identification information for \total \lirich stars in our sample. \galah IDs are \texttt{sobject\_id} in \cite{galah_sven_2021}, and \galex IDs are \texttt{objid} in \cite{galex_ais}. The full table in machine-readable format can be found online. \label{tb:full_rich}}
\tablehead{\colhead{\galah ID} & \colhead{\gaia DR2 ID} &\colhead{\gaia DR3 ID} & \colhead{\galex ID} } 
\startdata
131118002901313 & 4769316162914833024 & 4769316162914833024 & GALEX J052639.7-570922 \\
131120002001376 & 4690838726645395840 & 4690838726645395840 &  \\
131123003501064 & 5477688834192901632 & 5477688834192901632 &  \\
131218002401174 & 5479312744147026688 & 5479312744147026688 &  \\
140112002301046 & 5487706347195367168 & 5487706347195367168 &  \\
140209002201006 & 5295277759201045120 & 5295277759201045120 &  \\
140209002202072 & 5295162821578679168 & 5295162821578679168 &  \\
140303000402167 & 5390773299010265856 & 5390773299010265856 &  \\
140307003101263 & 6083719092007736704 & 6083719092007736704 &  \\
140309003101259 & 5246870110524338304 & 5246870110524338304 &  \\
... &... &... & ... \\
\enddata
\end{deluxetable*}
\begin{deluxetable*}{rrrr}
\tablecaption{Similar to Table \ref{tb:full_rich}. Survey identification information for \dgs in our sample. The full table in machine-readable format can be found online.\label{tb:full_poor}}
\tablehead{\colhead{\galah ID} & \colhead{\gaia DR2 ID} &\colhead{\gaia DR3 ID} & \colhead{\galex ID} } 
\startdata
131118002401169 & 4775062726077208704 & 4775062726077208704 & GALEX J043150.3-574434 \\
131118003401190 & 5494730267989420544 & 5494730267989420544 & GALEX J060423.2-585904 \\
131119001201296 & 5482949653733175936 & 5482949653733175936 & GALEX J062008.5-584639 \\
131120003501360 & 5478578579617154048 & 5478578579617154048 &  \\
131121001901108 & 5480375628293980928 & 5480375628293980928 &  \\
131121001901229 & 5483711821450447744 & 5483711821450447744 &  \\
131216002101129 & 5292810558187091328 & 5292810558187091328 &  \\
131216002101134 & 5293575577761817856 & 5293575577761817856 &  \\
140116004302034 & 5637744054654312960 & 5637744054654312960 & GALEX J091336.5-271813 \\
140305002601331 & 5673683104033521024 & 5673683104033521024 & GALEX J100455.8-163328 \\
... &... &... &... \\
\enddata
\end{deluxetable*}



\bibliography{references}{}
\bibliographystyle{aasjournal}

\end{document}